\documentclass[]{aa}  
\usepackage{graphicx}
\usepackage{amsmath}
\usepackage{natbib}
\usepackage{txfonts}
\bibpunct{(}{)}{;}{a}{}{,}
\begin{document} 

\title{Near-IR period-luminosity relations for pulsating stars in $\omega$~Centauri (NGC\,5139)\thanks{Based on observations collected at the European Organisation for Astronomical Research in the Southern Hemisphere, Chile, with the VISTA telescope (project ID 087.D-0472, PI R. Angeloni).}} 
   \author{C. Navarrete\inst{\ref{inst1}, \ref{inst2}} \and
          M. Catelan\inst{\ref{inst1}, \ref{inst2}, \ref{inst3}} \and
          R. Contreras Ramos\inst{\ref{inst2}, \ref{inst1}} \and
          J. Alonso-Garc\'ia\inst{\ref{inst4}, \ref{inst2}, \ref{inst1}} \and\\
          F. Gran\inst{\ref{inst1}, \ref{inst2}} \and
          I. D\'ek\'any\inst{\ref{inst5}} \and 
          D. Minniti\inst{\ref{inst6}, \ref{inst2}, \ref{inst7}}
          }         

   \institute{Instituto de Astrof\'isica, Pontificia Universidad Cat\'olica de Chile, Av. Vicu\~na Mackenna 4860, 782-0436 Macul, Santiago, Chile\\
              \email{cnavarre;mcatelan;rcontre@astro.puc.cl} \label{inst1} \and
              Instituto Milenio de Astrof\'isica, Santiago, Chile\label{inst2} \and
              Centro de Astro-Ingenier\'ia, Pontificia Universidad Cat\'olica de Chile, Santiago, Chile\label{inst3} 
              \and
              Unidad de Astronom\'ia, Facultad Cs. B\'asicas, Universidad de Antofagasta, Av. U. de Antofagasta 02800, Antofagasta, Chile	\label{inst4}
              \and
              Astronomisches Rechen-Institut, Zentrum f{\"u}r Astronomie der Universit{\"a}t Heidelberg, M{\"o}nchhofstr. 12-14, 69120 Heidelberg, Germany \label{inst5}
              \and
              Departamento de Fisica, Facultad de Ciencias Exactas, Universidad Andres Bello, Av. Fernandez Concha 700, Las Condes, Santiago, Chile \label{inst6}
              \and 
              Vatican Observatory, V00120 Vatican City State, Italy \label{inst7}
             }


  \abstract
   {}
   {$\omega$~Centauri (NGC~5139) hosts hundreds of pulsating variable stars of 
different types, thus representing a treasure trove for studies of their 
corresponding period-luminosity (PL) relations. Our goal in this study is to 
obtain the PL relations for RR Lyrae, and SX Phoenicis stars 
in the field of the cluster, based on high-quality, well-sampled light curves in 
the near-infrared (IR).}
   {$\omega$~Centauri was observed using VIRCAM mounted on VISTA. A total of 42 
epochs in $J$ and 100 epochs in $K_{\rm S}$ were obtained, spanning 352~days. 
Point-spread function photometry was performed using DoPhot and DAOPHOT in the 
outer and inner regions of the cluster, respectively.}
   {Based on the comprehensive catalogue of near-IR light curves thus secured, 
PL relations were obtained for the different types of pulsators in the cluster, 
both in the $J$ and $K_{\rm S}$ bands. This includes the first PL relations in 
the near-IR for fundamental-mode SX Phoenicis stars. The near-IR magnitudes and 
periods of Type II Cepheids and RR Lyrae stars were used to derive an updated 
true distance modulus to the cluster, with a resulting value of $(m-M)_0 = 
13.708 \pm 0.035 \pm 0.10$~mag, where the error bars correspond to the adopted  
statistical and systematic errors, respectively. Adding the errors in 
quadrature, this is equivalent to a heliocentric distance of $5.52\pm 0.27$~kpc.}
   {}
   

  \titlerunning{Near-IR PL relations for pulsating stars in $\omega$~Cen}
  \authorrunning{C. Navarrete et al.}
   \maketitle

\section{Introduction}\label{Intro}

$\omega$~Centauri (NGC~5139, C1323-472) is the most luminous, massive, and 
biggest globular cluster (GC) in the Milky Way. Located at RA~=~13:26:47.28 and 
DEC~=~-47:28:46.1 (J2000), it appears visible with the naked eye from the 
southern celestial hemisphere. It has been extensively studied, since it is 
bright ($M_V \sim -10.26$~mag), large (apparent size of $36\arcmin$) and nearby, 
at a distance of only 5.2~kpc, according to the 2010 version of the 
\citet{WH96,WH10} catalogue of GC parameters. Additionally, it hosts millions of 
stars in a field covering $\sim 1.5 \times 1.5  \, {\rm deg}^2$, which has led 
to numerous works devoted to its internal dynamics \citep{V06, V10, S13}, a 
suggested extragalactic origin, and its possible associated tidal debris 
\citep[e.g.][]{M01, D02, A05, M05, B06, DaCosta08, V11, FT15, CN15}. Naturally, 
it has also been the subject of extensive variability studies, which is also the 
main subject of the present work. 

In this context, $\omega$~Cen stands out as one of the three most RR Lyrae 
(RRL)-rich GCs known \citep{N15}. Moreover, it is the GC with the highest known 
number of SX Phe stars  \citep[74;][]{O05, W07, CS12}. At the same time, while 
the number of type II Cepheids (T2Cs) that it is known to harbor (seven) may 
seem small in an absolute sense, $\omega$~Cen still holds the record as the most 
T2C-rich of any GC \citep{C01, M06}. Also, even though anomalous Cepheids 
(ACEPs) are primarily found in nearby extragalactic environments \citep{CS15}, 
an ACEP classification has been advanced for some of the RRL stars in 
$\omega$~Cen \citep{N94}. The existence of ACEPs in $\omega$~Cen would thus be 
another indication of its possible extragalactic origin. However, none of the 
ACEP candidates has been confirmed or definitely rejected yet, probably because 
they are quite similar to the longest-period RRc stars in terms of periods and 
light curve shapes.

$\omega$~Cen is well known for hosting at least two stellar populations with 
metallicity peaks centered at ${\rm [Fe/H]} \approx -1.6$ and $-1.1$~dex 
\citep{B04, J13}, with additional peaks likely also being present 
\citep[e.g.,][]{CJ09}. Besides metallicity, it has been reported that these 
populations differ in their detailed chemistry and age, with the metal-richer 
stars likely being enhanced in helium by $\Delta Y \sim 0.17$ \citep[][and 
references therein]{N04, P05, D11, Dupree13, MT16}. Despite the strong evidence 
for large He-abundance spread among $\omega$~Cen's stars, \cite{S06} did not 
detect variations in $Y$ in their study of the pulsation properties (and 
spectroscopic metallicities) of the RRL stars in the cluster. Later, 
\cite{Marconi11} studied the impact of the helium content on the RRL properties 
based on evolutionary and pulsational models, finding that the helium content 
has a marginal effect on the pulsation properties of these variable stars. Very 
recently, \citet{MT16} claimed consistency between the predicted and observed 
properties of RRL stars in $\omega$~Cen but with a relatively small spread in 
helium, amounting to $\Delta Y \sim 0.03$. 

In this context, the different, well-represented variability types that are 
simultaneously present in $\omega$~Cen offer us a unique opportunity to perform 
a comparison of their properties. In particular, the pulsating stars present in 
the cluster are well-known distance indicators, following period-luminosity (PL) 
relations that are often used to derive distances \citep[][and references 
therein]{CS15}. Note, in this sense, that PL relations at infrared wavelengths 
have some advantages, compared to the ones derived at visible bands  
\citep[first highlighted by][in a study of the PL relations of Classical 
Cepheids]{M83}: the amplitude of the variables is smaller, leading to a smaller 
error in the derived mean magnitudes, even when relatively few data points are 
available; the interstellar extinction is lower ($\mathcal{A}_K/\mathcal{A}_V 
\sim 1/10$); and the infrared luminosities are less sensitive to temperature 
changes, leading to tighter PL relations, compared to the visible regime.

Among the pulsating variability types present in $\omega$~Cen, the primary 
distance indicators are the RRLs, which are not only relatively bright (being 
horizontal-branch [HB] stars) but also significantly more numerous than the 
other types of pulsators that are present in the cluster. RRL stars follow a 
well-defined PL relation at the near-IR, at odds with the visual where such a 
relation is not present \citep[see e.g.][]{CS15}. The near-IR PL relations 
were first discovered in the pioneering work by \cite{L86}, based on the study 
of three globular clusters. A detailed theoretical derivation in the 
Johnson-Cousin-Glass $UBVRIJHK$ system was performed by \cite{C04}, who showed 
how the PL relation first appears in the $R$ band, decreasing its scatter 
towards redder bands, becoming tightest in the $K$ regime. From the 
observational side, \cite{K14} studied RRLs from the field in 13 different 
photometric bandpasses, including the mid-infrared WISE bands, confirming that 
the PL relation does become tighter in the infrared \citep[see also][]{CK14}. 

SX Phe stars are the faintest among the known pulsating stars in the cluster, 
and thus more challenging to observe than their brighter siblings~-- and yet, 
they are the second most numerous type of variable stars currently known in 
$\omega$~Cen. The study of the PL relations for SX Phe and their Population~I 
counterparts, the $\delta$~Scuti stars, is complicated by the need to properly 
identify the pulsation modes, as these stars are often pulsating in more than 
one mode simultaneously. In these cases, the dominant mode is often the radial 
fundamental, first, or second overtone, though non-radial $p$ modes are often 
also present \citep[][and references therein]{CS15}. The PL relations of these 
stars have been investigated using theoretical models \citep[e.g.,][]{PCD99,T02} 
and observational data in the optical \citep{N94, MSX95, A00, M11, CS12} alike, but a 
clear consensus has not yet been reached regarding, for instance, the impact of 
the metallicity on these PL relations. An extension of these PL studies to the 
near-IR regime has not been carried out yet, and in fact is expected to present 
some important challenges: since many SX Phe stars are small-amplitude pulsators 
(with amplitudes below 0.2~mag or so in $V$), their amplitudes in the near-IR 
could easily become similar to, or even lower than, the photometric errors at 
their corresponding magnitudes. Such small amplitudes are detrimental to the 
proper characterization of the (often multiple) pulsation modes of the star, as 
needed to cleanly establish their PL relations. On the other hand, the smaller 
amplitudes make it easier to compute representative average magnitudes and 
colors from the time-series data. 

T2Cs are considerably fewer in number than either RRL or SX Phe, and only seven 
are known in $\omega$~Cen. Despite their brightness, the T2C PL relation has not 
been as extensively studied, undoubtedly due to the fact that they are so few in 
number \citep[see][for a review and extensive references]{CS15}. As in the case 
of the RRL stars, the PL relations for T2Cs have also been shown to become 
tighter at longer wavelengths \citep[][and references therein]{M06}. In the case 
of ACs, which are even fewer in number in GCs (but not in nearby dwarf galaxies, 
where they can be very numerous), PL relations in different bandpasses, 
including the near-IR, have recently been provided by \citet{VR14}. 

Our team has performed a systematic monitoring, in the $J$ and $K_{\rm S}$ 
bands, of $\omega$~Cen, with the main goal of obtaining a large number of 
well-defined light curves for the different types of variables present in this 
cluster. These well-defined light curves can play an important role as templates 
for the automated classification of variable stars in near-IR surveys 
\citep{A14}, including, in particular, the VISTA Variables in the V\'ia L\'actea 
\citep[VVV; e.g.,][]{M10, C11, S12DR1, C14s, C13} survey and the VISTA survey of the 
Magellanic Clouds system \citep[VMC; e.g.,][]{MRC11,VR14,VR16} both of which are 
time-resolved ESO Public Surveys conducted in the near-IR that run on ESO's 
VISTA telescope. 

This paper is the second in a series dedicated to presenting the results of our 
$\omega$~Cen near-IR variability survey. In the first paper, \cite{N15} studied 
the RRL population (in terms of variability type, amplitudes, and membership). 
Here we present the near-IR PL relations for the different types of pulsating 
variable stars that are present in the cluster. In the next and final paper of 
the series, Navarrete et al. (in prep.) will present the derived near-IR 
photometric catalogue for all detected variables in $\omega$~Cen, including not 
only the pulsating stars but also eclipsing binaries, ellipsoidal variables, 
etc. 

This paper is organized as follows. The observations and data reduction are 
presented in the next section. Section~\ref{sec:results} is devoted to the PL 
relations for the different types of pulsating stars in $\omega$~Cen. The 
implied pulsational distance modulus of the cluster and the sources of 
uncertainties are discussed in Section~\ref{sec:pdm}. Conclusions are presented 
in Sect.~\ref{conclusions}.


 \section{Observations}\label{Observations}
 
The VIRCAM camera \citep{D06}, mounted on the VISTA telescope \citep{E10}, was 
used to monitor $\omega$~Cen, obtaining 42 and 100 epochs in the $J$ and $K_{\rm 
S}$ bands, respectively. The effective field of view (FoV) of VISTA ($1.1 \times 
1.5 \, {\rm deg}^2$) is large enough to encompass all the pulsating stars known 
in the field of the cluster, except for four RRL located farther away than 
cluster's tidal radius and which are thus likely non-members \citep{N15}. 

The characteristics of the observations and data reduction are the same as those 
already explained in \cite{N13, N15}, and will accordingly not be repeated here. 
Point-spread function (PSF) photometry was performed using the photometry 
packages DoPhot \citep{S93,A12} for the outer regions, and DAOPHOT~II/ALLFRAME 
\citep{S87,Stetson94} for the central part of the cluster (i.e., the innermost 
$\sim 10\arcmin$). All the magnitudes are presented in the VISTA photometric 
system.  
 

\section{Results}\label{sec:results} 

From our time-series photometry, light curves with the clear signatures of 
variability were derived for 209 pulsating stars, including seven T2Cs, 89 
fundamental-mode (ab-type) and 98 first-overtone (c-type) RRL, 12 SX Phe, and three  
stars of uncertain type. Intensity-averaged magnitudes were obtained for all of 
these stars, based on Fourier fits to their light curves. For an additional 62 
SX Phe and four RRc stars, variability was not reliably recovered on the basis of 
our data alone, most likely due to these stars' very small amplitudes in the 
near-IR \citep[they were previously classified as variables by][based on 
optical photometry]{K04}. In these cases, the mean magnitudes were obtained 
directly from the photometric processing of the images, without any attempt to 
phase-fit the light curves. 

In order to compare the magnitudes and periods with well-calibrated near-IR PL 
relations, both $J$ and $K_{\rm S}$ magnitudes were dereddened, adopting a color 
excess of $E(B-V)= 0.12$~mag from the \citep{WH96,WH10} online catalogue, and 
the ratios $\mathcal{A}_X/E(B-V)$ of 0.866 and 0.364 for the VISTA $J$ and 
$K_{\rm S}$ magnitudes, respectively \citep{C11}. 

\subsection{Type II Cepheids} 

Light curves for all the T2Cs present in the cluster were recovered, both in $J$ 
and $K_{\rm S}$. However, the RV Tau star V1 has a magnitude near the saturation 
limit in the detector where it was observed. In particular, to derive its light 
curve, aperture photometry measurements (provided by the CASU catalogues) 
instead of PSF photometry were used because in the latter V1 is considered a 
saturated star. Because of this, the individual magnitude measurements have 
photometric errors higher than for the other T2Cs, and the errors adopted for 
the intensity-averaged magnitude, in $J$ and $K_{\rm S}$, are twice as high as 
the error adopted for the intensity-average magnitudes of the other T2Cs.

\begin{figure}
\centering
\includegraphics[width=0.5\textwidth]{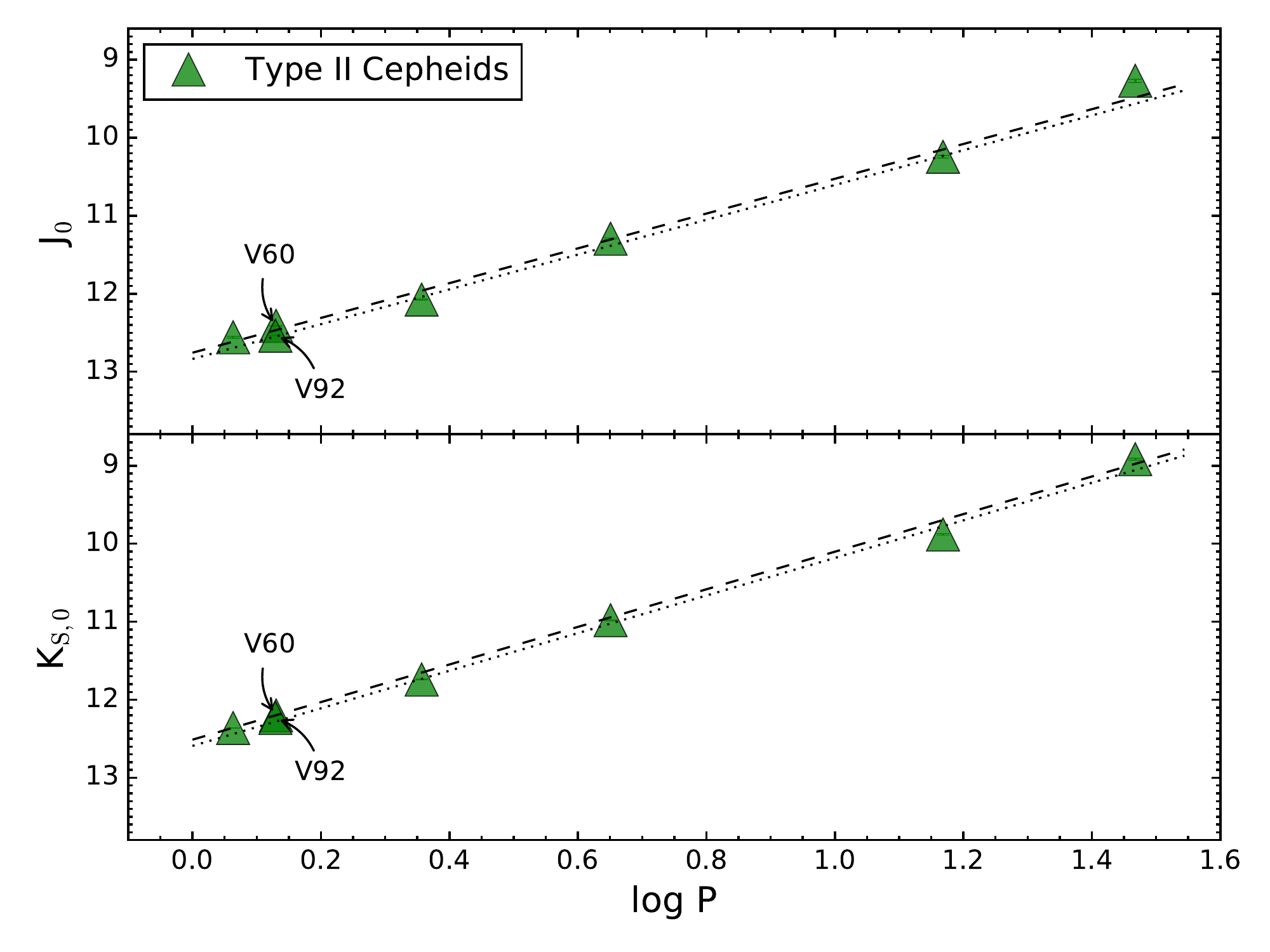}
\caption{Dereddened $J$ and $K_{\rm S}$ magnitudes versus period (in units of 
days) for the seven known T2Cs in $\omega$~Cen. The dashed lines correspond to 
the empirical PL relations derived by \cite{M06}, shifted to the distance 
modulus of the cluster used in that study, 13.62 mag, while the dotted lines are 
the same relations, but shifted to our derived distance modulus of 13.708 mag 
(see Section~\ref{sec:pdm}). V60 and V92, with roughly the same period and 
magnitude, are labeled accordingly.} \label{plt2cs}
\end{figure}

Figure~\ref{plt2cs} shows the intensity-averaged magnitudes versus $\log{P}$ for 
the seven T2Cs in $\omega$~Cen, for the $J$ and $K_{\rm S}$ bands. Dashed and 
dotted lines show the empirical PL relation derived by \cite{M06}, shifted by 
13.62 and 13.708 mag, respectively. The former value corresponds to the distance 
modulus adopted by \citeauthor{M06} for $\omega$~Cen, based on their adopted 
relation between HB luminosity and metallicity, while the latter one is the 
final distance modulus obtained in this study (see Section~\ref{sec:pdm}). As 
can be noted, all the variables closely follow a well-defined linear relation, 
irrespective of the distance modulus adopted, showing very little scatter (with 
a rms in the residuals of the weighted least-squares fit of 0.07 and 0.03~mag in 
the $J$ and $K_{\rm S}$ bands, respectively)~-- which is consistent with all of 
them pulsating in the same pulsation mode, and with a negligible dependence of 
their PL relation on metallicity (and helium). Indeed, \cite{M06} found that all 
T2Cs in Galactic GCs pulsate in the fundamental mode, and a similar conclusion 
has been reached by \cite{R14b} in a study of 130 T2Cs in the Large Magellanic 
Cloud. \citeauthor{M06} also found that the metallicity dependence of the PL 
relation is not very significant, with a zero point dependence of only $(-0.1\pm 
0.06)$~mag/dex in the $K_{\rm S}$ band. Based on theoretical models for BL Her 
stars (the short-period T2Cs), \cite{D07} similarly found a dependence of only 
$\sim 0.04$ and 0.06~mag~dex$^{-1}$ in $J$ and $K_{\rm S}$, respectively.   

While Figure~\ref{plt2cs} is consistent with all T2Cs pulsating in the 
fundamental mode, the previous study of \cite{N94} claimed that V1, V43, and V60 
are actually first-overtone pulsators, while V29, V48, V61, and V92 pulsate in 
the fundamental mode. However, Figure~\ref{plt2cs} does not show parallel 
sequences occupied by the different stars, as would be expected if they pulsated 
in different modes \citep{CS15}. In fact, the pair represented by V60 and V92 
overlap in Figure~\ref{plt2cs}, the former being brighter than the latter by 
only 0.15 and 0.05~mag in $J$ and $K_{\rm S}$, respectively.

Interestingly, V43 and V60 appear consistently brighter than the other stars at 
the same period (by 0.079 and 0.052~mag in $J$ and $K_{\rm S}$, respectively) 
when the PL relation of \cite{M06} is adopted, shifted to the distance modulus 
derived with the fundamental-mode candidates. We note that V1 deviates more 
dramatically from the \citeauthor{M06} relation, being brighter by $\sim 
0.25$~mag in $J$ and $\sim0.12$~mag in $K_{\rm S}$~-- but this is likely to be, 
at least in part, due to saturation (see above). Whether the measured shifts for 
V43 and V60 would be consistent with the stars pulsating in the first overtone 
should be further investigated by theoretical modelling of the PL relation for 
these stars. 

Adopting the calibrated PL relation derived by \cite{M06}, the distances for the 
seven T2Cs can be obtained from their intensity-averaged dereddened $J$ and 
$K_{\rm S}$ magnitudes. Table~\ref{table:t2c_dm} shows the true distance modulus 
of the cluster, determined using the weighted average of individual measurements 
in each bandpass, in three cases: (1) using all the T2Cs of the cluster; (2) 
discarding V1 (the only RV Tau star) due to the aforementioned saturation 
effects; (3) considering only the fundamental-mode candidates according to 
\cite{N94} (i.e., V29, V48, V61 and V92). Reassuringly, and as can be seen from 
the table, the final derived distance modulus does not depend on either the 
selected sub-sample or the adopted bandpass.  

\begin{table*}[ht!]
\centering
\caption{$\omega$~Cen weighted-average distance modulus, based on T2Cs.}\label{table:t2c_dm}
\begin{tabular}[h]{lccc}
\hline 
Sub-sample                       & $N_{\star}$  & $(J-M_J)_0$       & ($K_{\rm S}-M_{K_{\rm S}})_0$  \\
                                 &              & (mag)             & (mag)                          \\
\hline \hline
T2Cs (all)			 & 7            &  13.659 $\pm$ 0.121 & 13.646 $\pm$ 0.066               \\
T2Cs (all minus V1)              & 6            &  13.663 $\pm$ 0.073 & 13.646 $\pm$ 0.061               \\
Fundamental-mode candidates only & 4            &  13.686 $\pm$ 0.049 & 13.664 $\pm$ 0.061               \\
\hline
\end{tabular}
\end{table*}

\subsection{RR Lyrae stars}

\begin{figure}[t]
\centering
\includegraphics[width=0.5\textwidth]{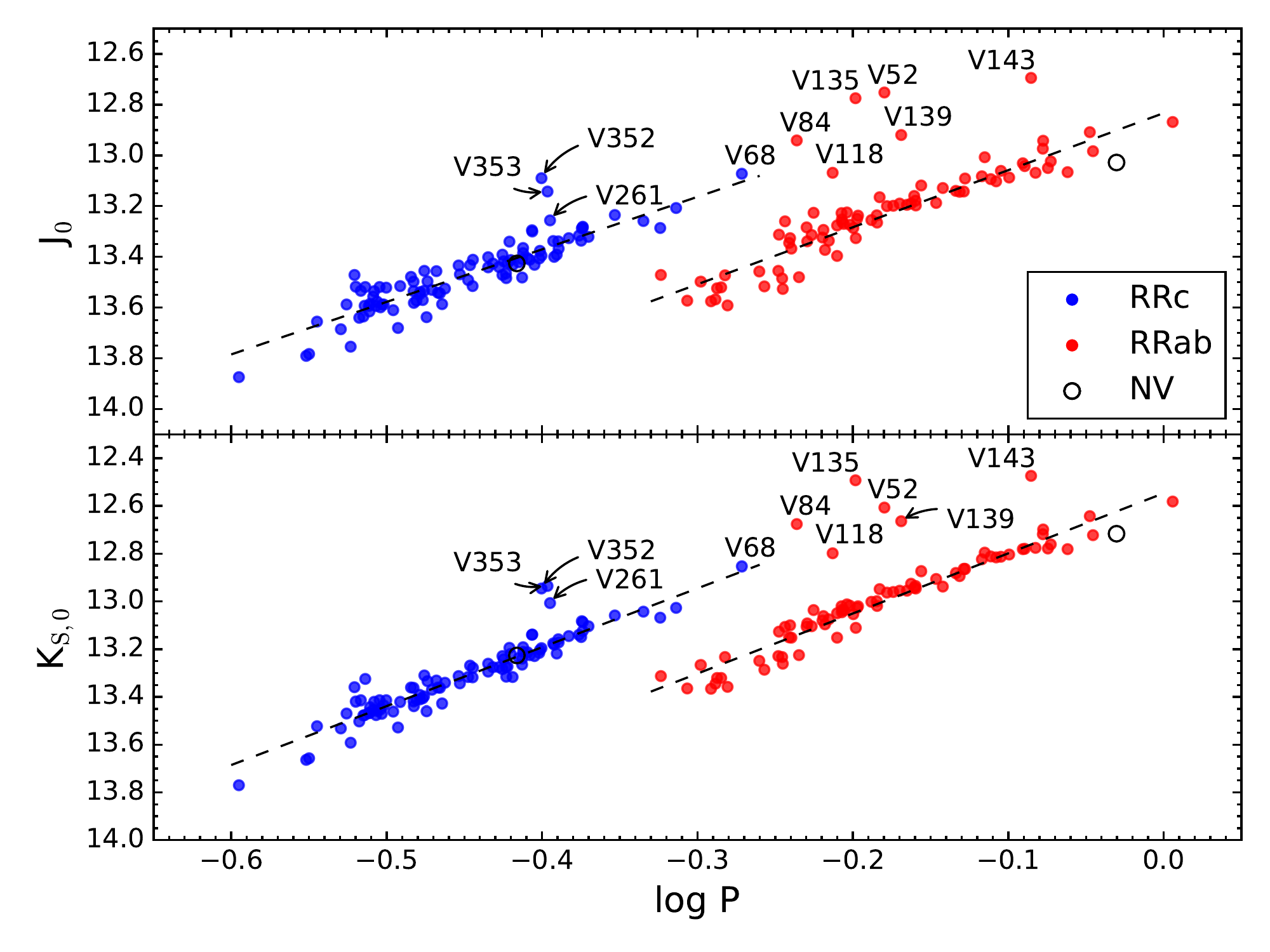}
\caption[Dereddened $J$ and $K_{\rm S}$ magnitudes versus period for RRLs in 
$\omega$~Cen]{Dereddened $J$ and $K_{\rm S}$ magnitudes versus period for RRLs 
in $\omega$~Cen. RRab stars are shown in red, and RRc stars are shown in blue. 
The dashed lines correspond to the best least-squares fits to the measurements, 
carried out separately for the ab- and c-type RRL, once the labeled stars have 
been removed from the samples (see text).}\label{plrrl}
\end{figure}

In order to conduct the most accurate study of the RRL PL relation, we first 
cleaned the sample, which contains 191 RRLs with measured $J$ and $K_{\rm S}$ 
magnitudes, by applying the following cuts. First, the field RRL stars V168, 
V181, V183, and V283 from \cite{K04}, and NV457 and NV458 from \cite{N15} were 
excluded, since they are not members of the cluster \citep{N15}. Next, V165 and 
V366 were also dropped, as they have periods of $\sim 0.5$ and 1.0~day, 
respectively, which does not allow the recovery of the full light curves. 
Finally, the RRc stars V349 and V351, which are located in the innermost region 
of the cluster and have inaccurate magnitudes, were discarded as well. As a 
result, a final sample of 83 RRab and 100 RRc stars was considered in our study of 
the PL relations in $J$ and $K_{\rm S}$. Their positions in the corresponding 
$\log{P}$-magnitude diagrams are shown in Figure~\ref{plrrl}. 


The four brightest RRL stars with periods $\sim$ 0.63~days ($\log{P} \sim$ 
--0.2), namely V84, V118, V135 and V139, have been referred to as 
``overluminous'' previously by \cite{DP06}, but from the examination of the 
\textit{HST} catalogue of \cite{A10}, all but V84 have very near companions 
which could lead to unresolved blends \citep[see][]{N15}. V52 and V143 were not included 
in the study of \cite{DP06} but were found to be brighter than other RRab stars 
because of marked blends \citep{N15}. From this point onwards, these five RRab 
stars will be discarded from our analysis. V84 is an almost isolated star and 
its brightness appears to be intrinsic, which could be because it is a 
foreground RRL or, as \cite{N94} commented, an ACEP candidate. This possibility 
will be addressed in Section~\ref{acs} and, for now, the star will also be 
discarded from the RRL sample. 

In Figure~\ref{plrrl} there are three RRc stars that appear a little above the 
other RRc stars with the same period. They are V261, V352 and V353, all with 
periods of about $\sim 0.4$~days ($\log{P}\sim -0.4$). If their magnitudes are 
well constrained, then their periods might be consistent with second-overtone 
pulsations, in which case they would be candidate RRe stars, for which one might 
expect $P_2/P_0 \approx 0.57$ \citep[e.g.,][]{CS15}. However, RRe stars are 
normally expected to be bluer than the RRc stars and to have short periods 
\citep[][and references therein]{CS15}, whereas the long-period V261 and V353 
appear redder instead, and fall near the RRab stars in the CMD \citep{N15}. V352 
appears brighter than the RRc stars, but not bluer. It is perhaps more likely 
that these stars are affected by blends, and should accordingly be discarded 
from our analysis. Note that, as these three stars do not have previous 
metallicity measurements, they were not considered in the derivation of empirical 
PL-Z relations, neither in the distance modulus calculation.

Table~\ref{tab:rrlyrae_met} lists all the cluster member RRL stars with 
metallicity measurements from \cite{R00} and/or \cite{S06}, including their 
equatorial coordinates, periods, $J$ and $K_{\rm S}$ intensity-averaged mean 
magnitudes and their corresponding (statistical) errors, subtypes and 
metallicities. Since the number of RRL stars with spectroscopic metallicity is 
large, our sample allows us to obtain purely empirical PL-$Z$ relations both in 
$J$ and $K_{\rm S}$.  To do this, we adopted the metallicities from \cite{S06} 
and ``fundamentalized'' the periods of RRc stars, using the relation $\log{P_0} 
= \log{P_1} + 0.127$. We prefer the metallicities from \cite{S06} instead of 
those derived by \cite{R00} because the former were derived based on 
high-resolution spectra. The metallicities were transformed into $\log{Z}$ using 
the relation
\begin{equation}
 \log{Z} = {\rm [Fe/H]}+\log{\left(0.362+0.638 \, f\right)}+ \log{Z_{\odot}}
 \label{eq:z}
\end{equation}

\begin{figure}[ht!]
 \centering
 \includegraphics[width=0.5\textwidth]{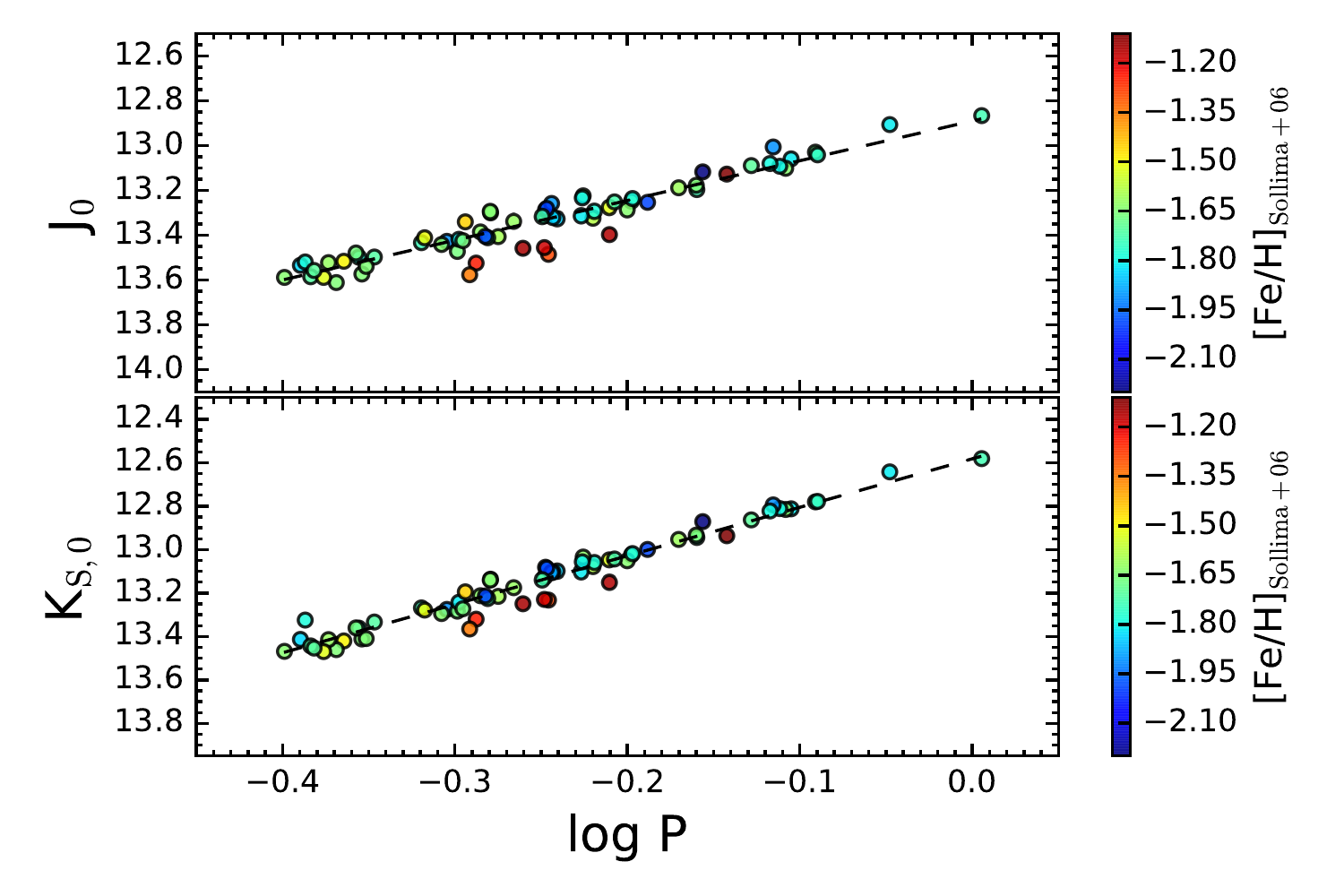}
  \caption[Dereddened $J$ and $K_{\rm S}$ magnitudes versus period considering 
metallicities]{{\em Top panel}: Dereddened $J$ magnitudes versus period with 
metallicities from \cite{S06} in the color-bar. {\em Bottom panel}: The same as in 
the top panel, but for the $K_{\rm S}$ magnitude. Dashed lines are the 
empirical PL-$Z$ relations derived from our data set, evaluated at the mean RR 
Lyrae [Fe/H] = --1.67 dex.}\label{plfeh}
\end{figure}

\noindent \citep{Salaris93} where the $\alpha$-element enhancement factor $f = 
10^{\rm [\alpha/Fe]}$. For consistency with \cite{C04}, $Z_{\odot}$~=~0.017 and 
$f = 3$ were adopted. The coefficients and one standard deviation errors of the 
least-square fit are reported in Table~\ref{table:coef}. Figure~\ref{plfeh} 
displays the sample used and the PL-$Z$ relations obtained both for $J$ and 
$K_{\rm S}$, evaluated at the mean metallicity of the RRLs, [Fe/H] = --1.67 dex. 
As can be seen from the plot, the $J$-band relation presents more scatter than 
the $K_{\rm S}$ one, which is consistent with theoretical expectations 
\citep[e.g.,][]{C04} and with the fact that the $J$-band light curves present 
higher amplitudes, which can induce more scatter in the corresponding average 
quantities. Note that the value of the $J$-band $\log{P}$ slope $a$ is in 
excellent agreement with the theoretical predictions, whereas there is a small 
disagreement in the case of $K_{\rm S}$, with the predicted value of $a$ being 
larger (in absolute value) by about 0.1 \citep[see Table~3 in][]{Coppola11}. As 
far as the metallicity dependence is concerned, our derived $b$ slopes are 
significantly steeper than was reported in \cite{Sollima06b}, but only slightly 
steeper than predicted by the \cite{C04} models. Naturally, we anticipate that 
the Gaia mission \citep{C16} will provide the accurate distances to RR Lyrae 
stars that are needed in order to establish the definitive slopes and zero 
points of the PL-$Z$ relations. In the meantime, however, considering the 
reasonable agreement between the \cite{C04} model predictions and our results, 
we will employ the former in the remainder of our analysis.

\begin{table}[h!]
\centering
\caption{Empirical PL-$Z$ relations for RRL stars in $\omega$~Cen. The relation 
has the form $m_X = a\log{P}+b{\rm [Fe/H]}+c$, where $X$ corresponds to the bandpass.}\label{table:coef}
\begin{tabular}{ccccc}
\hline
Band               & $a$    & $b$ & $c$ & $R^2$\\
\hline \hline 
\multicolumn{5}{c}{$m_X = a\log{P}+b{\rm [Fe/H]}+c$} \\
\hline
$J$         & --1.774 $\pm$ 0.061 & 0.153 $\pm$ 0.027 & 13.079 $\pm$ 0.075 & 0.936 \\
$K_{\rm S}$ & --2.232 $\pm$ 0.044 & 0.141 $\pm$ 0.020 & 12.752 $\pm$ 0.054 & 0.985 \\
\hline
\end{tabular}
\end{table}

In order to estimate the distance modulus of the cluster based on RRL stars, the 
calibrated theoretical PL-$Z$ relations in the VISTA filter system for the $J$ and 
$K_{\rm S}$ bands \citep[eqs. 1 and 3 in][which were adapted from Catelan et 
al. 2004]{A15} were used. PL relations for RRL stars have a non-negligible 
dependence on metallicity, even though the dependence is lower in the near-IR 
bands compared to the optical \citep[e.g.,][]{C04}. Using the metallicities 
spectroscopically derived for RRL stars in $\omega$~Cen by \cite{R00} and 
\cite{S06}, the distance modulus for the cluster was calculated using: (1) only 
the RRab stars; (2) only the RRc stars; and (3) all RRab and RRc stars. The 
periods of RRc stars were fundamentalized using the relation $\log{P_0} = 
\log{P_1} + 0.127$ \citep[corresponding to a period ratio $P_1/P_0 = 0.746$, as 
adopted by][]{DP06}. [Fe/H] values for 64 and 53 RRab and RRc stars, 
respectively, are available in the catalogue of \cite{R00}, while from 
\cite{S06} the metallicities of 33 RRab and also 31 RRc stars were considered. 
Seven RRL from \citeauthor{R00} and four from \citeauthor{S06} were not 
considered as they do not have reported errors in the metallicities, being those 
measurements highly uncertain. Both metallicity catalogues were compared by 
\citeauthor{S06}, who found a systematic offset of 
$\Delta$[Fe/H](\citeauthor{S06}$-$\citeauthor{R00}) $= -0.06$~dex, with a 
dispersion of 0.3~dex. Because of that, the two metallicity sources were treated 
separately. The weighted-average distance modulus found with the $J$ and $K_{\rm 
S}$ magnitudes of the individual RRLs in the three different cases are listed in 
Table~\ref{table:rrl_dm}.

\begin{table*}[h!]
\centering
\caption{Distance moduli based on RRLs using spectroscopic metallicities and different pulsational modes.}\label{table:rrl_dm}
\begin{tabular}[h!]{lcc}
\hline 
                       & ($J-M_{\rm J})_0$  & ($K_{\rm S}-M_{\rm K_{\rm S}})_0$   \\
                       & (mag)              &  (mag) \\
\hline \hline
\multicolumn{3}{c}{{\bf Metallicities from Rey et al. (2000)}} \\
\hline
RRab (64 stars)      &  13.728 $\pm$ 0.063 & 13.752 $\pm$ 0.043 \\
RRc  (53 stars)      &  13.685 $\pm$ 0.058 & 13.709 $\pm$ 0.047 \\
RRab+RRc (117 stars) &  13.698 $\pm$ 0.064 & 13.722 $\pm$ 0.049 \\
\hline
\multicolumn{3}{c}{{\bf Metallicities from Sollima et al. (2006)}}\\
\hline
RRab (33 stars)      &  13.706 $\pm$ 0.042 & 13.746 $\pm$ 0.027 \\
RRc  (31 stars)      &  13.693 $\pm$ 0.047 & 13.728 $\pm$ 0.040 \\
RRab+RRc (64 stars)  &  13.698 $\pm$ 0.046 & 13.735 $\pm$ 0.036 \\
\hline
\end{tabular}
\end{table*}

\begin{figure}
 \centering
 \includegraphics[scale=0.45]{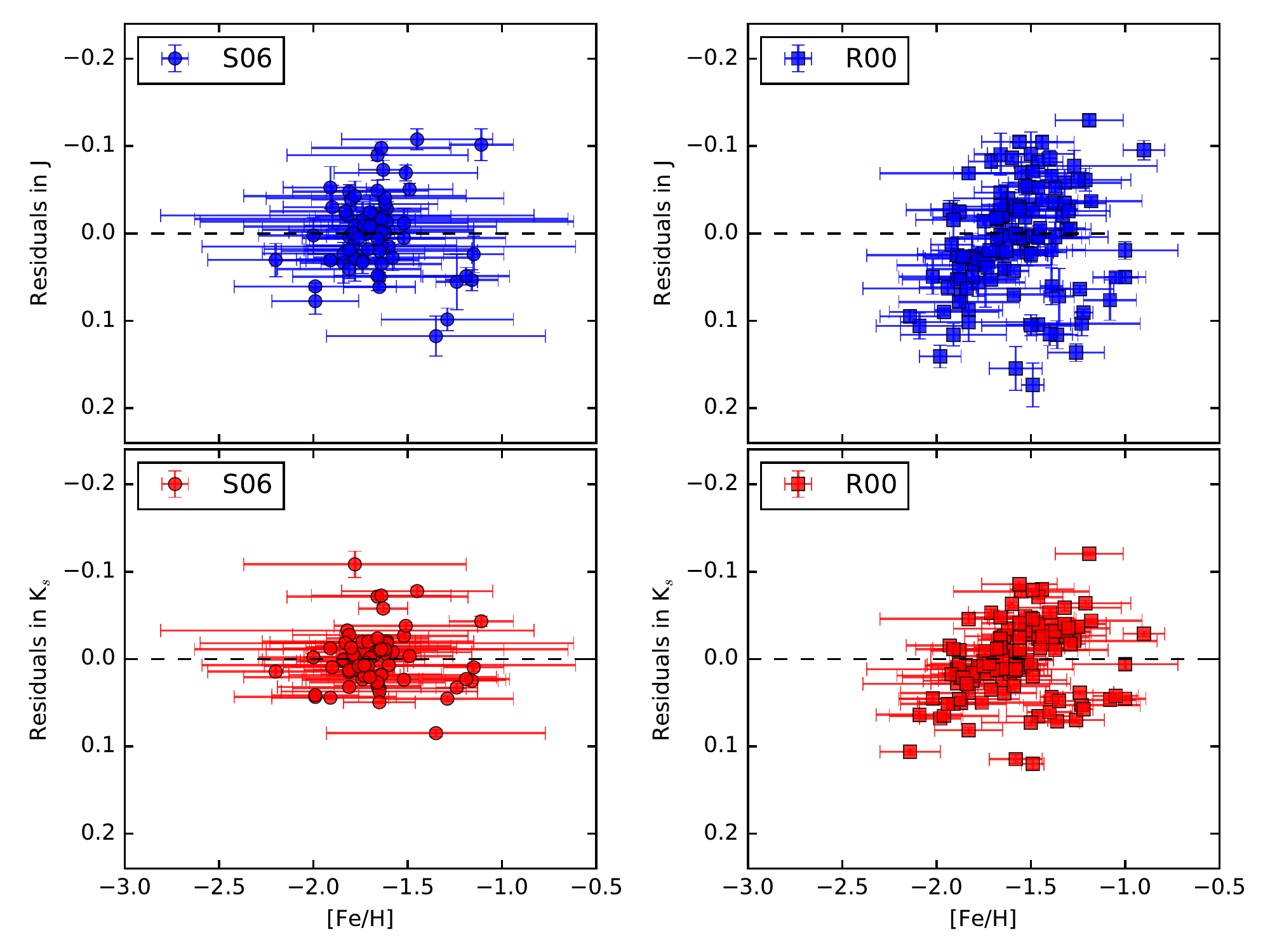}
 \caption{Difference between the observed mean magnitudes and the derived PL-$Z$ 
relations as a function of metallicity. Top panels show the residuals for $J$, 
while the residuals in $K_{\rm S}$ are shown in the bottom panels. Left and 
right panels correspond to the metallicities derived by \cite{S06} and 
\cite{R00}, respectively. \label{fig:residuals}}
\end{figure}

Figure~\ref{fig:residuals} shows the magnitude differences between the mean 
$J$ (upper panels) and $K_{\rm S}$ (bottom panels) magnitudes and the derived 
PL-$Z$ relations (listed in Table~\ref{plfeh}), as a function of the metallicity 
from \cite{S06} and \cite{R00}. As expected, there is less scatter for the 
$K_{\rm S}$ magnitudes (right panels), compared to the residuals for the $J$ 
band (left panels). The 3$\sigma$ residuals are $\sim$0.13 and 0.10 mag for $J$ 
and $K_{\rm S}$, respectively, considering the metallicities derived by 
\cite{S06}. When the metallicities of \cite{R00} are adopted, the 3$\sigma$ 
residuals are slightly larger, with 0.19 and 0.13 mag for $J$ and $K_{\rm S}$ 
magnitudes, respectively. For both $J$ and $K_{\rm S}$, the residuals as a 
function of the [Fe/H] values derived by \cite{S06} are clustered around zero, 
and there is no evident trend with metallicity, proving that the metallicity 
term in the derived PL-$Z$ relations is well represented by the fit. 
Nonetheless, for the metallicities derived by \cite{R00}, there seems to be a 
small correlation between metallicities and magnitudes. However, we emphasize 
that our relations (Table~\ref{table:coef}) were derived using the metallicities 
from \cite{S06}, who used high-resolution spectroscopy in their work, presumably 
leading to more accurate values than in the case of 
the $\Delta$S method used by \cite{R00}.

\subsection{Anomalous Cepheids}
\label{acs}

Figure~\ref{plrrl} shows that one RRab star, namely V84, is brighter than any 
other RRab that follows the PL relation, reaching the region where the RRc stars 
are located. However, the position in the CMD is in agreement with a cluster 
member star \citep{N15}. The period was first derived by \cite{B02} and 
confirmed by the $J$ and $K_{\rm S}$ light curves, discarding an RRc 
classification. However, \cite{S90}, \cite{L90}, and \cite{N94} all proposed 
that V84 and also V15, V68 (the longest-period RRc star in the cluster) and V99 
could be ACEPs, rather than RRL stars. 

Sparse measurements in the near-IR have been done for ACEPs. Only recently, 
\cite{R14} presented a large catalogue of ACEPs in the Large Magellanic Cloud 
which have been gathered using near-IR observations performed by the VISTA 
Magellanic Survey \citep[VMC;][]{C11VMC}. 

\begin{figure}[t]
\centering
\includegraphics[width=0.5\textwidth]{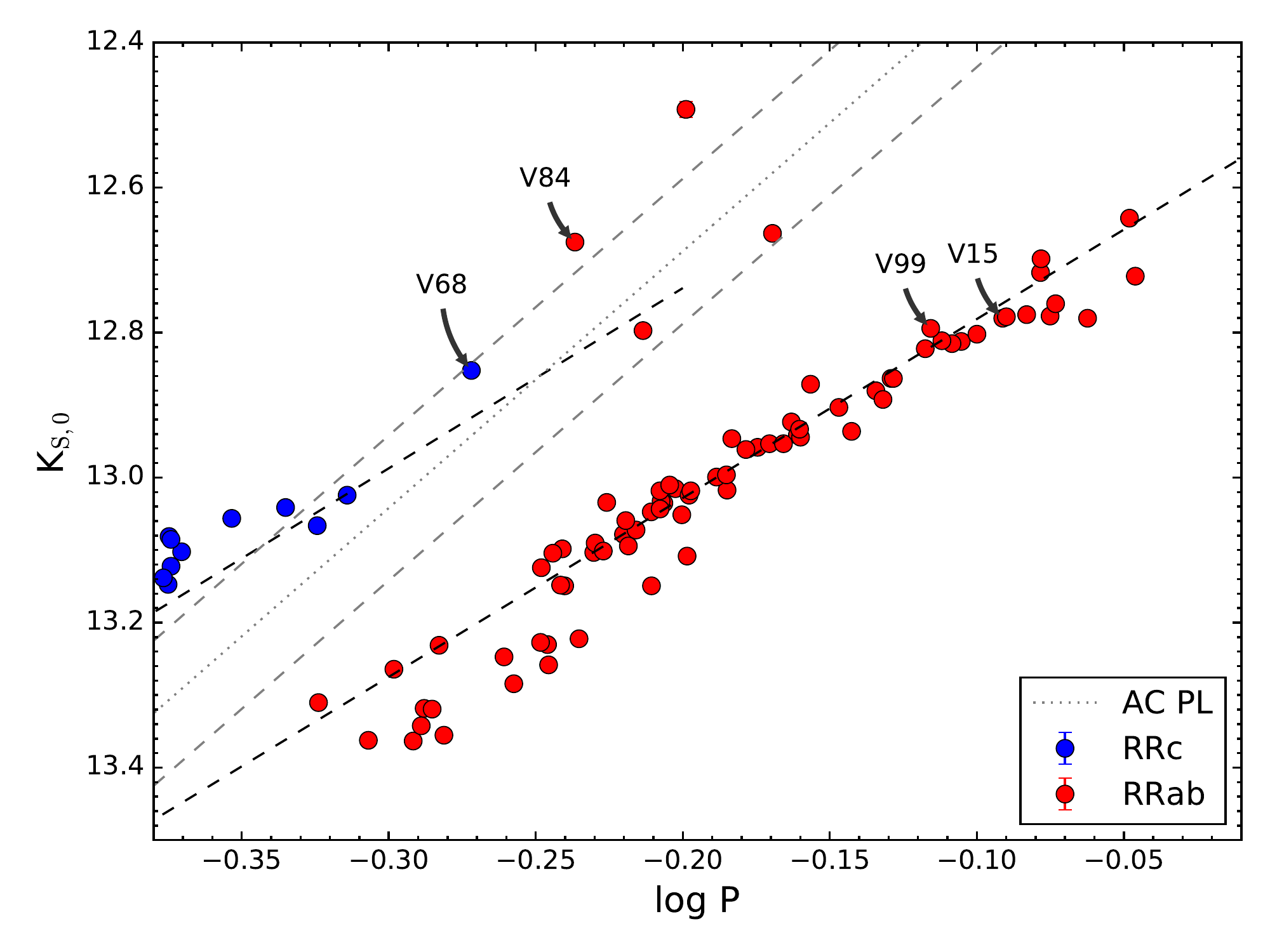}
\caption[V68 and V84 as possible ACEPs]{Dereddened $K_{\rm S}$ magnitudes versus 
period around the locations of V68 and V84, two candidate ACEPs in $\omega$~Cen. 
The mean trend of RRab and RRc are denoted by the black dashed lines, whereas 
the dotted line shows the $K_{\rm S}$-PL relation for ACEPs, as derived by 
\cite{R14}, shifted to the adopted distance modulus of $\mu_0$~=~13.708 mag. The 
gray dashed lines represent $\pm0.1$~mag deviations around the latter 
relation.}\label{plac}
\end{figure}

Figure~\ref{plac} shows the PL relation for the ACEPs at the $\omega$~Cen 
distance using the calibrated PL relation derived by \cite{R14}, and adopting 
the $\omega$~Cen's true distance modulus (see Section~\ref{sec:pdm}) of 
$\mu_0$~=~13.708 mag. The PL relation is shown as a dotted line and the dashed 
lines above and below it represent $\pm0.1$~mag deviations.

As can be noted, between $\log{P} \sim -0.35$ and $\sim -0.18$, the ACEPs and 
RRc PL relations cover a similar region of magnitudes, which does not allow to 
cleanly separate both types of stars based on this diagram, if they are both at 
the same distance. Thus, using the magnitudes from VISTA it is not possible to 
confirm nor to reject the possibility that V68 and V84 are ACEPs in 
$\omega$~Cen. If V84 is not an ACEP, it should be a foreground RRab. V15 and 
V99, the other RRab stars proposed as ACEP candidates, follow closely the RRab 
PL relation, as can also be seen in Figure~\ref{plfeh}, discarding the 
possibility of being ACEP variables.

\subsection{SX Phoenicis}

The $\omega$~Cen field contains a large number of SX Phoenicis stars~-- larger, 
in fact, than for any other GC~-- thus providing an excellent opportunity to 
calibrate the SX Phe PL relation. Previous such studies have however been 
limited to the optical. 

\cite{N94} considered the $B$ and $V$ magnitudes of three SX Phe stars from 
$\omega$~Cen as well as 11 more from other GCs in order to derive observational 
PL relations. \cite{M11} considered most of the known SX Phe in $\omega$~Cen 
\citep[not including the five SX Phe stars from][]{W07}, finding a high spread 
in the $\log{P}-\left<V\right>$ diagram (see his Fig.~9), which may be partially 
explained by the variation of metallicity values in the cluster stars. 
Nevertheless, there are no previous metallicity measurements for the SX Phe 
stars in the cluster. According to \cite{M11}, the scatter can be reduced by 
selecting a subsample of SX Phe stars with long periods ($\log{P} \geq -1.47$), 
which he argues are likely to be metal-poor (${\rm [Fe/H]} \leq -1.0$), 
fundamental-mode pulsators. 

\begin{figure}[t]
 \centering
 \includegraphics[width=0.5\textwidth]{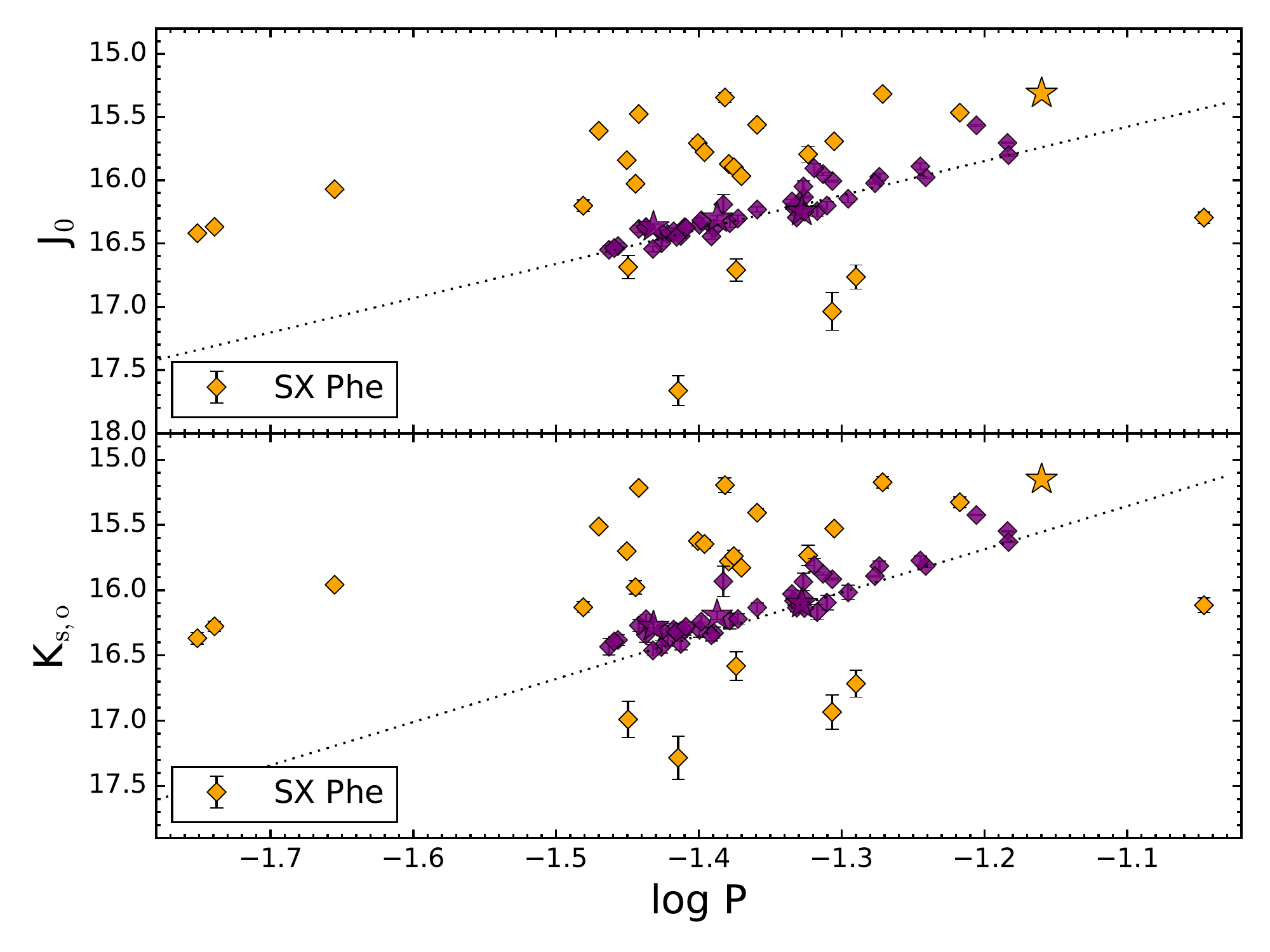}
 \caption[Dereddened $J$ and $K_{\rm S}$ magnitudes versus period for the SX Phe 
stars]{Dereddened $J$ and $K_{\rm S}$ magnitudes versus period for all the SX 
Phe stars that appear to be members of the cluster. Those from \cite{K04} are 
plotted as orange diamonds while the star symbols denote those discovered by 
\cite{W07}. The ``lower-envelope'' stars that tentatively follow a common trend 
and would correspond to fundamental-mode pulsators are shown in purple.}
 \label{sx_all}
\end{figure}

When plotting the $\log{P}-J$ and $-K_{\rm S}$ diagrams, the same situation as 
in \citet{M11} (i.e., high spread) is found, as can be see in 
Figure~\ref{sx_all}. The field SX Phe stars, namely the foreground star V65 and 
the background stars V297 and ID-92 from \cite{W07}, were not included in the 
diagram. 

The large amount of scatter notwithstanding, Figure~\ref{sx_all} also reveals 
the presence of a tight ``lower envelope'' of stars, below which only a handful 
of faint stragglers are found. We conjecture that this may correspond to the 
sequence of fundamental-mode pulsators in the cluster. The five stars located 
below this lower envelope (V302, V307, V315, V323 and V324) are probably 
non-radial pulsators. The three stars with the shortest-periods (V294, V295 and 
V296, with $\log{P} < -1.6$) are likely high-overtone pulsators. The 
longest-period star, V328, is considered a fundamental-mode pulsator 
\citep{O05}, but its magnitudes in the visible and in the $J$ and $K_{\rm S}$ 
bands are fainter than the linear trend that the other stars at shorter periods 
define. Because of that, it will also be excluded to derive the fundamental-mode 
PL relation for SX Phe stars. 

\begin{figure}[t]
 \centering
 \includegraphics[width=0.5\textwidth]{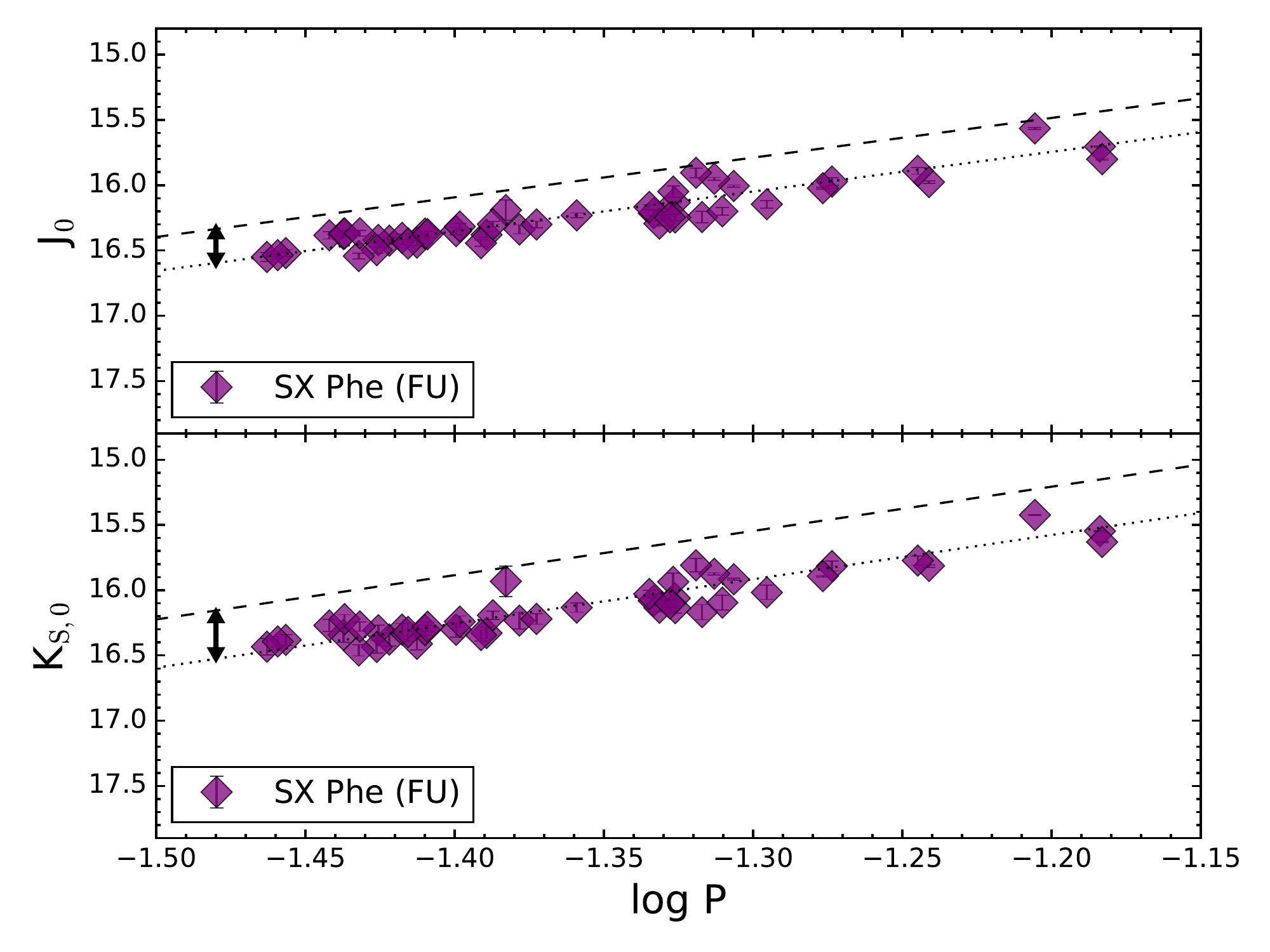}
 \caption[Fundamental-mode SX Phe candidates]{Fundamental-mode SX Phe candidates 
from which observational $J_{\rm 0}$ and $K_{\rm S,0}$ PL relations were 
derived. The dotted lines correspond to the best least-squares fit associated to 
each filter. The dashed lines corresponds to the expected position of the 
first-overtone pulsators, assuming that fundamental and first-overtone PL 
relations are parallel.}
 \label{sx_fm}
\end{figure}

Table~\ref{tab:sx} presents the variable ID, equatorial coordinates, 
periods, intensity-averaged $J$ and $K_{\rm S}$ magnitudes for the 45 SX 
Phe fundamental-mode candidates. The stars that exhibit light curves with clear 
evidence of variability are explicitly tagged as ``Variability recovered'' in 
this table. Figure~\ref{sx_fm} shows the thus selected 45 fundamental-mode 
candidates, where there of the four SX Phe stars from \cite{W07} are included. In 
Figure~\ref{sx_fm}, the dotted lines represent the least-squares best fit to the 
fundamental-mode candidates, while the dashed lines are the expected PL relation 
for first-overtone pulsators, assuming that both relations are parallel and 
adopting a ratio of the first-overtone period to the fundamental-mode period of 
$\sim$~0.775 \citep{M11}, which implies a magnitude shift (at fixed $\log{P}$) 
of 0.26 and 0.37~mag for $J$ and $K_{\rm S}$, respectively. 

Based on these 45 selected SX Phe stars, observational PL relations for the 
fundamental mode candidates obtained are

\begin{align}
J_0 &= -(3.04 \pm 0.17) \log{P} + (12.10 \pm 0.22), \label{eq1} \\
&\sigma = 0.09 \text{ mag,}\nonumber  \\
K_{\rm S,0} &= -(3.39 \pm 0.24) \log{P} + (11.51 \pm 0.30)\rm{,} \label{eq2} \\ 
& \sigma = 0.11 \text{ mag.}\nonumber 
\end{align}

The dispersion around both these relations are quite similar and smaller than 
that associated to the $V$-band relation derived by \cite{M11}, being the latter 
0.13~mag. This difference is mainly due to the intrinsic dispersion of the 
optical PL relations compared to the near-IR, for any type of pulsating variable 
star \citep{CS15} Despite the fact that most of the light curves of the SX Phe 
stars could not be fully recovered, their behavior in the $\log{P}-$magnitude 
diagram is consistent with most of them pulsating in the fundamental mode.

\begin{figure}[t]
 \centering
 \includegraphics[width=0.5\textwidth]{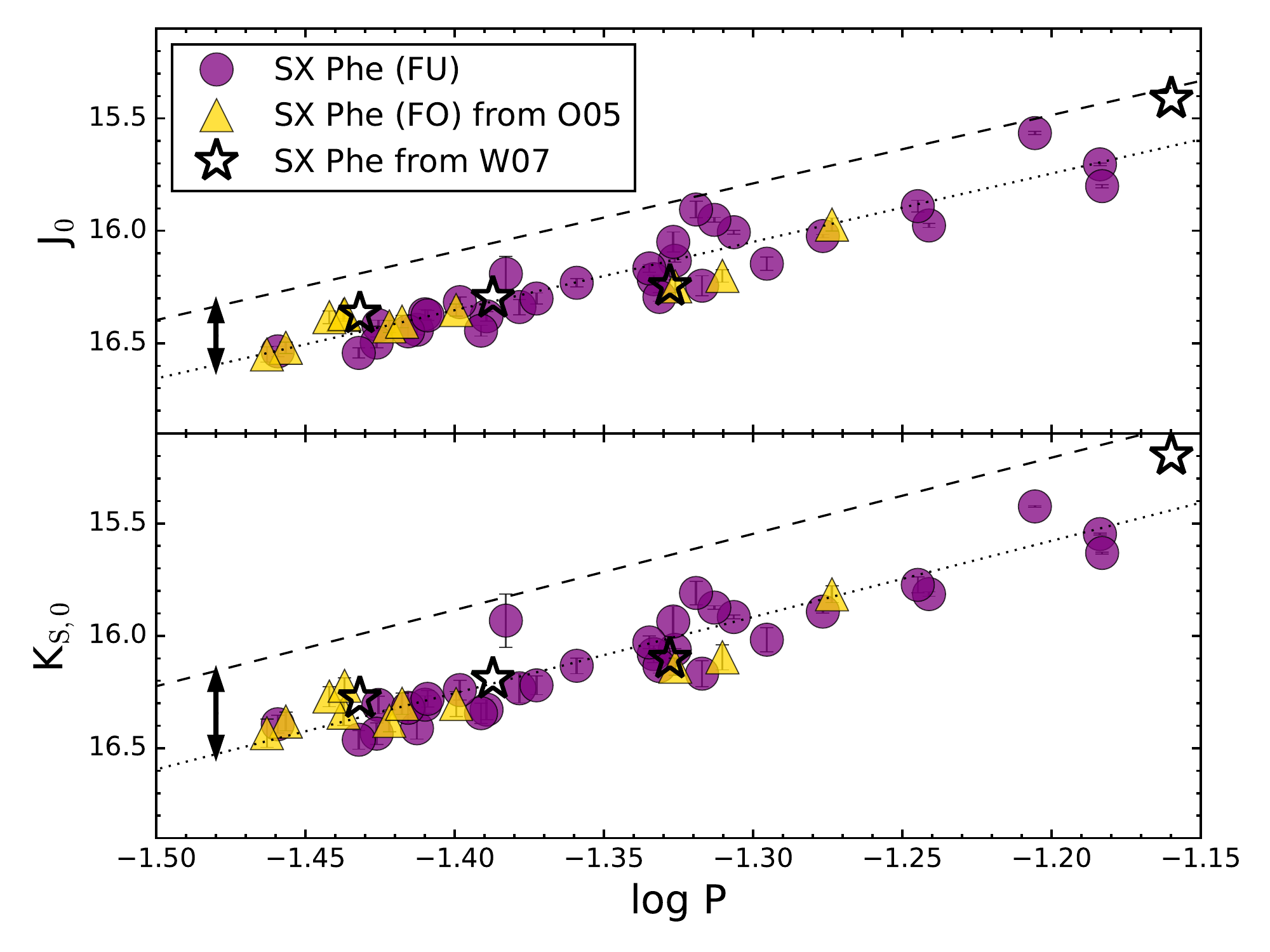}
  \caption[Fundamental and first-overtone mode SX Phe from 
\cite{O05}]{Fundamental (in purple circles) and first-overtone (in  
yellow triangles) SX Phe as classified by \cite{O05}, based on their frequency 
spectrum. The dotted lines are the PL relations derived previously. As can be 
seen, there are some claimed first-overtone pulsators that also follow the 
relations defined by fundamental-mode pulsators.}
 \label{sx_fm_olech}
\end{figure}

However, according to \cite{O05}, there are some first-overtone SX Phe among  
the ones that we have considering as fundamental-mode pulsators. 
Figure~\ref{sx_fm_olech} shows the same stars as Figure~\ref{sx_fm} but now the 
fundamental-mode and first-overtone candidates as derived by \citeauthor{O05} 
are plotted as purple circles and yellow triangles, respectively. The diagram includes 
the four SX Phe (members of the cluster) discovered by \citeauthor{W07}, 
after the study of \citeauthor{O05} was published (open stars).

From the figure, the SX Phe found by \citet{W07} are consistent with three of 
them pulsating in the fundamental mode. On the other hand, the one with the 
longest period, namely ID-7 (at $\log{P} \sim -1.16$), appears as a bona fide 
first-overtone pulsator. The other SX Phe stars, despite being considered 
first-overtone pulsators by \cite{O05}, follow the PL relations as given by 
equations~\ref{eq1} and~\ref{eq2}, suggesting that they too may be 
fundamental-mode pulsators. 

\section{On the $\omega$~Cen pulsational distance modulus}\label{sec:pdm}

\subsection{Systematic errors}

Before we adopt a pulsational distance modulus for $\omega$~Cen, based on the 
different values obtained using T2Cs and RRL stars, the different sources of 
systematic errors will be evaluated.

In the case of T2Cs, one possible source of systematic error in the distance 
modulus values is that some first overtone pulsators could be considered as 
pulsating in the fundamental mode. To estimate the error associated to this, we 
could consider the difference between the distance modulus obtained when 
different stars are used, as listed in Table~\ref{table:t2c_dm}. Adopting the 
$J$ and $K_{\rm S}$-based true distance moduli derived using W Vir and BL Her 
stars, the systematic errors should be at least 0.04 and 0.03~mag, respectively. 
The metallicity term could constitute an additional source of systematic errors 
but, given the low metallicity dependence of the T2C PL relation, we expect that 
these errors are lower than 0.01~mag, both in $J$ and in $K_{\rm S}$.

In the case of RRL stars, one source of systematic error could be provided by 
the sample choice, i.e., including or not fundamental and first-overtone 
pulsators simultaneously in the fits. \cite{DP06} found, based on synthetic HB 
models,  that the relations for fundamental and first-overtone RRL are not 
exactly parallel, and that the use of fundamentalized RRc periods produces an 
increase (by a factor of 2) in the uncertainty of the zero points and slopes of 
their relations. Nevertheless, \cite{C04} still found tight near-IR PL 
relations, considering RRab and (fundamentalized) RRc at the same time. 

Table~\ref{table:rrl_dm} lists the three possible cases, i.e., using only RRab, 
only RRc (with fundamentalized periods) and RRL stars from both pulsational 
modes (again with RRc periods fundamentalized). The metallicities derived by 
\cite{S06} seem to be more accurate (i.e., close to the actual metallicity of 
the stars) than the ones derived by \citeauthor{R00}, in the sense that more 
precise distance moduli are obtained. The fact that the results are always 
consistent (to within the errors), irrespective of the adopted subsample, 
suggests that RRab and (fundamentalized) RRc stars do indeed follow closely the 
same PL relation in the near-IR. 

Comparison between the different distance moduli obtained using the two 
different sources of metallicity in Table~\ref{table:rrl_dm} suggests to us that 
the metallicity scale can be one source of systematic error in the final derived 
distance modulus. Based on these results, we adopt a value of 0.02 and 
0.01~mag, for $J$ and $K_{\rm S}$ respectively, as representative of the main 
identified sources of systematic errors affecting the $\omega$~Cen distance 
modulus, as derived on the basis of the near-IR PL relations for RRL stars. 
Moreover, the individual metallicities for RRL stars have errors of the order of 
$\sim 0.19$ and $0.31$~dex, as reported by \cite{R00} and \cite{S06}, 
respectively. These uncertainties, propagated into the PL-Z relations used to 
derive the distance modulus, add the following errors: 0.04 and 0.03~mag for $J$ 
and $K_{\rm S}$ distance modulus, with metallicities from \citeauthor{R00}, and 
0.06 and 0.05~mag for the $J$ and $K_{\rm S}$ values derived with the 
metallicities from \citeauthor{S06}. At the same time, results from previous 
studies \citep[e.g.,][]{Cacciari03,DP06,F08,Gran15} suggest that distances derived 
on the basis of such relations can probably not yet be considered accurate to 
better than about $\pm0.1$~mag. Thus, we adopt the latter value as a 
conservative estimate of the systematic error affecting the distance moduli 
provided in our study.  

\subsection{Pulsational distance modulus from T2Cs and RRL stars}

Table~\ref{table:dm_both} lists the weighted-average distance modulus values 
from T2Cs and RRL stars and sigma as the dispersion around the mean. The weights 
for each individual distance modulus values were defined as the inverse of the 
squared of the total error, including statistical and systematic errors (summed up  
in quadrature). As expected, the error is lower for the $K_{\rm S}$-PL relations 
compared to $J$, since the former have lower intrinsic dispersion and is based 
on a more extensive dataset. Comparing both variability types, T2Cs lead to a 
less accurate distance modulus, probably because they are few in number, which 
increases the statistical error, and they have higher systematic errors since 
the effect of metallicity and pulsation mode separation in the derivation of the 
PL relations is not completely established.

Considering the distance modulus values found using T2Cs stars as well as RRL 
stars, both in $J$ and $K_{\rm S}$, the weighted-averaged distance modulus to 
$\omega$~Cen was determined, leading to a final value of $\mu_0 = 
13.708 \pm 0.035$~mag, where the error bar corresponds to the standard deviation 
of the mean. Given the discussion of systematic errors in the previous 
section, our final adopted distance modulus is 
$\mu_0 = 13.708 \pm 0.035 \pm 0.10 $~mag, where the error bars correspond to 
the adopted statistic and systematic errors, respectively. Adding the errors in 
quadrature \citep[as recommended, for instance, by][]{rbea02}, 
this corresponds to a heliocentric distance of $5.52\pm 0.27$~kpc.
 
As a further check, we have also explored the results of \citet{MF11}, who 
suggested that T2Cs and RRL follow basically the same PL relation in $K$. Using 
metallicities for the T2Cs from \citet{GW94} and \citet{vL07}, and assuming that 
the same near-IR relation as for the RRL applies also to the T2Cs, we obtain for 
the latter a distance modulus of $\mu_0 = 13.708 \pm 0.035$~mag (standard 
deviation of the mean), in excellent agreement with the previously derived 
value. On the other hand, carrying out the same analysis using the $J$-band data 
does not provide similarly consistent results, with a resulting distance modulus 
that is shorter by about 0.3~mag. This suggests that the T2Cs and RRL stars may 
indeed follow the same PL relation in $K_{\rm S}$, but not in $J$. 

Table~\ref{mu_lit} lists some of the true distance modulus values for 
$\omega$~Centauri found in the literature. In order to put all of them in the 
same system, the color excess was set as $E(B-V) = 0.12$~mag, a standard 
extinction law with $R_V = 3.1$ was adopted, and the apparent $V$ distance 
moduli were converted accordingly. The work of \cite{C02} listed some of 
$\omega$~Cen's distance moduli derived in the literature and compared them, 
considering the influence of the  metallicity spread as well as the possible 
helium enhancement associated to second-generation stars in the cluster. 
Distance modulus values based on PL-HB type relations were derived for different 
$\alpha$-enhancement levels by \cite{DP06}, and compared with other values from 
the literature.

The distance modulus found in this work appears in excellent agreement with most 
of the values listed in Table~\ref{mu_lit}. The distance modulus derived by 
\cite{DP06}, using a combination of near-IR light curves and 2MASS single-epoch 
magnitudes for different RRL stars in the cluster, appears quite similar to the 
average value found using our fully phase-folded light curves 
derived from VISTA filters.

\subsection{Difference between T2C- and RRL-based distance moduli}

It should be noticed that the $\omega$~Cen distance modulus values derived using 
T2Cs appear shorter than those found with the RRL stars, by an amount ranging 
from 0.03~mag in $J$ to 0.08~mag in $K_{\rm S}$ (see Table~\ref{table:dm_both}). 
Such a difference could not be explained by the metallicity term in the T2Cs PL 
relations as, if included, the corresponding distance moduli would be even 
lower. 

One possible explanation for the difference could rest on the multiple 
populations in $\omega$~Cen and the associated helium enhancement \citep{C02, 
S06, Marconi11}. He-rich stars are expected to populate the bluest part of the 
HB, where the T2C progenitors are thought to be located, thus those stars are 
affecting the magnitudes and pulsational periods of T2Cs with respect to the 
expected ones for stars with primordial helium abundance.

\begin{table}[t!]
\centering
\caption{Adopted distance moduli based on T2Cs and RRL stars.}\label{table:dm_both}
\begin{tabular}[h]{lcccc}
\hline 
          & $\mu_0(J)$ & $\sigma_{\rm total}$ & $\mu_0(K_{\rm S})$ & $\sigma_{\rm total}$ \\
          & (mag) & (mag) & (mag) & (mag) \\
\hline \hline
T2Cs      &  13.675 & 0.042 & 13.652 & 0.031 \\
RRLs      &  13.701 & 0.024 & 13.733 & 0.017 \\
\hline
\end{tabular}
\end{table}

\begin{table*}[ht!]
\centering
\setlength\tabcolsep{1.1ex}
\caption{Distance modulus for $\omega$~Cen from the literature}\label{mu_lit}
\begin{tabular}{lcl}
\hline
Method & $\mu_0$ (mag)$^{a}$ & Reference  \\ 
\hline \hline
$B$ and $V$ PL relations for Pop. II stars                          & $13.53 \pm 0.20$ & \cite{N94} \\
High-amplitude $\delta$ Sct stars                                   & $14.05 \pm 0.10$ & \cite{Mc00}$^{b}$ \\
Detached eclipsing binary (V212)                                    & $13.68 \pm 0.11$ & \cite{T01} \\
Detached eclipsing binary (V212)                                    & $13.72 \pm 0.04$ & \cite{K02}\\
$A_V$-$\log{P}$ models for RRab stars                               & $13.64 \pm 0.11$ & \cite{C02} \\
$M_V$-$\log{P}$ models for RRc stars                                & $13.71 \pm 0.11$ & \cite{C02} \\
$M_V-{\rm [Fe/H]}$ relation (RRL stars)                             & $13.72 \pm 0.11$ & \cite{DP06}$^{c}$ \\
$M_V-{\rm [Fe/H]}$ relation (RRL stars)                             & $13.62 \pm 0.11$ & \cite{DP06}$^{d}$ \\
PL-$Z$($K_{\rm S}$)                                                 & $13.77 \pm 0.07$ & \cite{DP06}$^{e}$ \\
PL-$\tau_{\rm HB}$($K_{\rm S}$, $f=3$, $\tau_{\rm HB} = 0.940$)     & $13.69 \pm 0.06$ & \cite{DP06} \\
PL-$\tau_{\rm HB}$($J$, $f$~=~3, $\tau_{\rm HB} = 0.940$)           & $13.70 \pm 0.10$ & \cite{DP06} \\
PL-$\tau_{\rm HB}$($K_{\rm S}$, $f=3$, $\tau_{\rm HB} = 0.934$)     & $13.72 \pm 0.06$ & \cite{DP06}$^{f}$ \\
PL-$\tau_{\rm HB}$($J$, $f$~=~3, $\tau_{\rm HB} = 0.934$)           & $13.76 \pm 0.10$ & \cite{DP06} \\
Cluster dynamics                                                    & $13.41 \pm 0.14$ & \cite{V06} \\
$M_V-{\rm [Fe/H]}$ relation (RRL stars)                             & $13.68 \pm 0.27$ & \cite{W07} \\
SX Phe $\log{P}-V$ relation                                         & $13.76 \pm 0.06$ & \cite{M11} \\
Optical Period-Wesenheit relations (RRL stars)                      & $13.71 \pm 0.08$ & \cite{B16} \\
\hline
Weighted average  & $13.72 \pm 0.13$ & \\
\hline
{\bf Weighted-average PL relations (T2Cs and RRLs)} & {\bf $13.708 \pm 0.035$} & {\bf This work} \\
\hline                                       
\end{tabular} \\
\begin{flushleft}
\footnotesize{$^{a}$ Most of the error estimates do not consider the 
contribution of systematic errors, which in the case of T2Cs and RRL stars can 
reach values of order $\approx 0.10$~mag.\\ $^{b}$ The listed error was 
determined considering the internal uncertainties of the mean distance modulus 
value, $\pm$0.02 mag, and the observational error, $\pm$0.1 mag, as described by 
\cite{Mc00}. \\ $^{c}$ Based on the $M_V$-[Fe/H] relation from \cite{B03} and 
mean $V$ magnitudes from \cite{K04}. \\ $^{d}$ Based on the $M_V$-[Fe/H] 
relation of \cite{C04} and $V$ magnitudes from \cite{K04}. \\ $^{e}$ Based on 
the semi-empirical $J$-band PLZ relation from \cite{B03}. \\ $^{f}$ Based on the 
PL-$\tau_{\rm HB}$ relations from \cite{C04}.}
\end{flushleft}
\end{table*}

Nonetheless, \cite{Marconi11} developed evolutionary and pulsational models in 
order to derive the number of variable stars in the HB of $\omega$~Cen 
considering different helium abundances. The authors compared the number 
fraction of T2Cs over the number of variable stars in the instability strip as 
predicted by the models and the observed one. For $Y =0.24$, 2\% of T2Cs are 
found with the models, in good agreement with the observational 3\%. Helium 
enhancement is taken into account considering two stellar populations with 
primordial initial helium abundances (80\% of the stars) and enhanced-helium ($Y 
=0.3$, 20\% of the stars). It was found that the percentage of expected T2Cs, 
5\%, would be higher than the one actually observed. This suggests that T2C 
stars in $\omega$~Cen are predominantly not He-enhanced.

For RRL stars, the authors found that the minimum fundamentalized period 
predicted by the models is in good agreement with the observed values of 
0.34~days when the primordial helium abundance is considered. At higher helium 
abundance values, the shortest fundamentalized period increases, which is not 
observed. This also suggests that RRL stars in $\omega$~Cen are not He-enhanced.

We thus conclude that other sources of systematic error will have to be 
investigated, in order to explain the origin of the difference in distance 
moduli obtained on the basis of the T2C and RRL near-IR PL relations.

\section{Conclusions}\label{conclusions}

We have presented the implications of our recent, extensive time-series photometry of $\omega$~Cen \citep{N15} for the near-IR (VISTA $J$ and $K_{\rm S}$) PL relations of different types of pulsators along the ``classical'' instability strip. These include T2Cs, ACEPs, RRL, and SX~Phe stars. For the T2Cs and RRL stars, for which the corresponding near-IR relations had previously been calibrated, we use the results to obtain a new distance modulus estimate for the cluster, $ \mu_0 = 13.708 \pm 0.035 \pm 0.10$~mag, where the error bars correspond to the adopted 
statistical and systematic errors, respectively. Adding the errors in quadrature, this is equivalent to a heliocentric distance of $5.52\pm 0.27$~kpc. This is in excellent agreement with previous results from the literature.\footnote{After we submmited our paper, \cite{Bhardwaj2017} published new near-IR PL and Period-Wesenheit relations for T2Cs, using template-fit $I$ and $K_{\rm S}$ light curves. Combining both relations, the authors derived a true distance modulus for $\omega$~Cen of $\mu(K_{\rm S}) = 13.70 \pm 0.11$~mag, in good agreement with our derived distance modulus based on RRL stars, but slightly different from the derived distance modulus using T2Cs and the $K$-band relation derived by \cite{M06}.} An offset between the T2C and RRL-based result, at the level of a few hundredths of a magnitude (with the T2Cs implying a smaller distance), is however present, for reasons which are not completely clear at present. 

The sizable number of RRL stars in the cluster allows us to derive new empirical 
near-IR PL-$Z$ relations in the $J$ and $K_{\rm S}$ bandpasses. Adopting as the 
true distance modulus of the cluster $\mu_0 = 13.708$~mag, the PL-$Z$ relations 
presented in Table~\ref{table:coef}, in terms of the absolute magnitude, are given by

\begin{align}
M_J({\rm RRL})           = -&(1.77 \pm 0.06) \log{P}  \nonumber \\
			    &+ (0.15 \pm 0.03) {\rm [Fe/H]} -(0.63 \pm 0.08), \label{eq_rrlj} \\
M_{K_{\rm S}}({\rm RRL}) = -&(2.23 \pm 0.04) \log{P}  \nonumber \\
			   &+ (0.14 \pm 0.02) {\rm [Fe/H]} -(0.96 \pm 0.05) \rm{.} \label{eq_rrlk}  
\end{align}

For the ACEPs, we show that their expected positions in the near-IR PL 
relations are very similar to those of RRc stars. This makes it difficult to 
properly establish their pulsation status on the basis of our near-IR data 
alone.

For the SX~Phe stars, we provide, for the first time, a calibration of 
their near-IR PL relation. Using a true distance modulus of $\mu_0 = 13.708$~mag for 
the cluster, and combining with equations~\ref{eq1} and \ref{eq2}, we obtain: 

\begin{align}
M_J({\rm SX~Phe})           &= -(3.04 \pm 0.17) \log{P} - (1.60 \pm 0.22), \label{eq1b} \\
M_{K_{\rm S}}({\rm SX~Phe}) &= -(3.39 \pm 0.24) \log{P} - (2.19 \pm 0.30)\rm{.} \label{eq2b}  
\end{align}

In the next (and final) paper of this series (Navarrete et al., in prep.), 
we will present our full near-IR catalogue of variable stars in the $\omega$~Cen 
field.  
  
\begin{acknowledgements}
We warmly thank the referee, B. Madore, for very useful suggestions that helped to improve our paper. Support for this project is provided by the Ministry for the Economy, Development, and Tourism's Millennium Science Initiative through grant IC\,120009, awarded to the Millennium Institute of Astrophysics (MAS); by the Basal Center for Astrophysics and Associated Technologies (CATA) through grant PFB-06/2007; by CONICYT's PCI program through grant DPI20140066; and by FONDECYT grants \#1141141 and \#1171273 (C.N., M.C.), \#1130196 (D.M.), \#1150345 (F.G.), \#3130320 (R.C.R.), and \#11150916 (J.A.-G.). C.N. acknowledges additional support from CONICYT-PCHA/Doctorado Nacional 2015-21151643.
\end{acknowledgements}

\bibliography{biblio}

\begin{appendix}
  \section{Catalogue}
  
\clearpage

\begin{table*}[ht!]
\centering
\caption[]{Catalogue of RR Lyrae stars in $\omega$~Cen field, used to derive the empirical PL relations) \label{tab:rrlyrae_met}}
\begin{tabular}{lcccccccccl}
\hline \hline
  ID   & RA (J2000.0)  & DEC (J2000.0) & $P$    & J     & e     & Ks    & e     & Type   & [Fe/H]         & [Fe/H]         \\
       & hh:mm:ss.ss    & dd:mm:ss.s     & (days) & (mag) & (mag) & (mag) & (mag) &      & (R00)          & (S06)          \\
\hline
 V3    & 13:25:56.14 & -47:25:54.2 & 0.841258   & 13.152 & 0.008 & 12.821 & 0.003 & RRab & -1.54$\pm$0.05 &                 \\                 
 V4    & 13:26:12.91 & -47:24:19.2 & 0.627320   & 13.376 & 0.051 & 13.059 & 0.005 & RRab & -1.74$\pm$0.05 &                 \\                 
 V5    & 13:26:18.31 & -47:23:12.8 & 0.515274   & 13.626 & 0.032 & 13.362 & 0.005 & RRab & -1.35$\pm$0.08 & -1.24$\pm$0.11  \\ 
 V7    & 13:27:01.02 & -47:14:00.1 & 0.713000   & 13.290 & 0.012 & 12.947 & 0.003 & RRab & -1.46$\pm$0.08 &                 \\                 
 V8    & 13:27:48.43 & -47:28:20.7 & 0.521329   & 13.575 & 0.013 & 13.275 & 0.004 & RRab & -1.91$\pm$0.28 &                 \\                 
 V9    & 13:25:59.57 & -47:26:24.4 & 0.523315   & 13.694 & 0.025 & 13.399 & 0.006 & RRab & -1.49$\pm$0.06 &                 \\ 
 V10   & 13:26:06.99 & -47:24:37.0 & 0.375125   & 13.493 & 0.007 & 13.271 & 0.003 & RRc  & -1.66$\pm$0.10 &                 \\                 
 V11   & 13:26:30.54 & -47:23:01.9 & 0.564798   & 13.415 & 0.008 & 13.168 & 0.004 & RRab & -1.67$\pm$0.13 & -1.61$\pm$0.22  \\                
 V12   & 13:26:27.17 & -47:24:06.6 & 0.386769   & 13.467 & 0.004 & 13.234 & 0.002 & RRc  & -1.53$\pm$0.14 &                 \\                 
 V14   & 13:25:59.66 & -47:39:09.9 & 0.377114   & 13.567 & 0.004 & 13.317 & 0.002 & RRc  & -1.71$\pm$0.13 &                 \\                 
 V15   & 13:26:27.08 & -47:24:38.4 & 0.810642   & 13.133 & 0.013 & 12.824 & 0.002 & RRab & -1.64$\pm$0.39 & -1.68$\pm$0.18  \\                
 V16   & 13:27:37.71 & -47:37:35.2 & 0.330202   & 13.675 & 0.005 & 13.453 & 0.002 & RRc  & -1.29$\pm$0.08 & -1.65$\pm$0.46  \\                
 V18   & 13:27:45.06 & -47:24:56.9 & 0.621689   & 13.372 & 0.014 & 13.079 & 0.003 & RRab & -1.78$\pm$0.28 &                 \\                 
 V19   & 13:27:30.12 & -47:28:05.7 & 0.299551   & 13.857 & 0.004 & 13.632 & 0.002 & RRc  & -1.22$\pm$0.05 &                 \\                 
 V20   & 13:27:14.03 & -47:28:06.9 & 0.615559   & 13.379 & 0.017 & 13.091 & 0.003 & RRab &                & -1.52$\pm$0.34  \\             
 V21   & 13:26:11.15 & -47:25:59.3 & 0.380812   & 13.535 & 0.011 & 13.358 & 0.003 & RRc  & -0.90$\pm$0.11 &                 \\                 
 V22   & 13:27:41.05 & -47:34:08.1 & 0.396127   & 13.508 & 0.005 & 13.257 & 0.002 & RRc  & -1.63$\pm$0.17 & -1.60$\pm$0.99  \\              
 V23   & 13:26:46.48 & -47:24:39.6 & 0.510870   & 13.678 & 0.023 & 13.407 & 0.004 & RRab & -1.08$\pm$0.14 & -1.35$\pm$0.58  \\              
 V24   & 13:27:38.33 & -47:34:15.0 & 0.462278   & 13.361 & 0.005 & 13.085 & 0.002 & RRc  & -1.86$\pm$0.03 &                 \\                 
 V25   & 13:26:25.49 & -47:28:23.7 & 0.588466   & 13.386 & 0.021 & 13.147 & 0.004 & RRab & -1.57$\pm$0.14 &                 \\                 
 V26   & 13:26:23.61 & -47:26:59.9 & 0.784720   & 13.163 & 0.009 & 12.856 & 0.002 & RRab & -1.68$\pm$0.10 & -1.81$\pm$0.12  \\       
 V27   & 13:26:26.04 & -47:28:16.0 & 0.615680   & 13.499 & 0.012 & 13.193 & 0.003 & RRab & -1.50$\pm$0.26 & -1.16$\pm$0.14  \\        
 V30   & 13:26:15.91 & -47:29:56.5 & 0.404410   & 13.440 & 0.005 & 13.217 & 0.002 & RRc  & -1.75$\pm$0.17 & -1.62$\pm$0.28  \\                 
 V32   & 13:27:03.34 & -47:21:39.2 & 0.620347   & 13.365 & 0.026 & 13.076 & 0.007 & RRab & -1.53$\pm$0.16 &                 \\                 
 V33   & 13:25:51.57 & -47:29:06.1 & 0.602324   & 13.426 & 0.015 & 13.122 & 0.004 & RRab & -2.09$\pm$0.23 & -1.58$\pm$0.42  \\            
 V35   & 13:26:53.24 & -47:22:34.9 & 0.386841   & 13.487 & 0.005 & 13.254 & 0.002 & RRc  & -1.56$\pm$0.08 & -1.63$\pm$0.36  \\               
 V36   & 13:27:10.19 & -47:15:29.5 & 0.379683   & 13.515 & 0.007 & 13.261 & 0.002 & RRc  & -1.49$\pm$0.23 &                 \\                 
 V38   & 13:27:03.23 & -47:36:30.4 & 0.779061   & 13.205 & 0.006 & 12.859 & 0.002 & RRab & -1.75$\pm$0.18 & -1.64$\pm$0.40  \\            
 V39   & 13:27:59.82 & -47:34:42.3 & 0.393374   & 13.533 & 0.004 & 13.271 & 0.002 & RRc  & -1.96$\pm$0.29 &                 \\                 
 V40   & 13:26:24.54 & -47:30:46.7 & 0.634072   & 13.352 & 0.025 & 13.068 & 0.004 & RRab & -1.60$\pm$0.08 & -1.62$\pm$0.19  \\                
 V41   & 13:27:01.37 & -47:31:02.0 & 0.662942   & 13.302 & 0.021 & 13.005 & 0.004 & RRab & -1.89$\pm$0.48 &                 \\                 
 V44   & 13:26:22.38 & -47:34:35.7 & 0.567545   & 13.587 & 0.013 & 13.274 & 0.003 & RRab & -1.40$\pm$0.12 & -1.29$\pm$0.35  \\               
 V45   & 13:25:30.85 & -47:27:20.9 & 0.589116   & 13.441 & 0.016 & 13.134 & 0.004 & RRab & -1.78$\pm$0.25 &                 \\ 
 V46   & 13:25:30.23 & -47:25:51.7 & 0.686971   & 13.292 & 0.008 & 12.967 & 0.002 & RRab & -1.88$\pm$0.17 &                \\                 
 V47   & 13:25:56.47 & -47:24:12.3 & 0.485123   & 13.309 & 0.010 & 13.068 & 0.006 & RRc  & -1.58$\pm$0.31 &                \\                 
 V49   & 13:26:07.73 & -47:37:55.9 & 0.604627   & 13.475 & 0.013 & 13.138 & 0.003 & RRab & -1.98$\pm$0.11 &                \\                 
 V50   & 13:25:53.92 & -47:27:36.2 & 0.386172   & 13.584 & 0.003 & 13.307 & 0.003 & RRc  & -1.59$\pm$0.19 &                \\                 
 V51   & 13:26:42.58 & -47:24:21.6 & 0.574152   & 13.429 & 0.023 & 13.142 & 0.004 & RRab & -1.64$\pm$0.21 & -1.84$\pm$0.23 \\                 
 V52$^{a}$ & 13:26:35.16 & -47:28:03.8 & 0.660386   & 12.854 & 0.022 & 12.650 & 0.007 & RRab & -1.42$\pm$0.04 &               \\
 V54   & 13:26:23.50 & -47:18:48.1 & 0.772915   & 13.196 & 0.008 & 12.855 & 0.003 & RRab & -1.66$\pm$0.12 & -1.80$\pm$0.23 \\                 
 V55   & 13:25:45.11 & -47:42:20.0 & 0.581724   & 13.582 & 0.014 & 13.266 & 0.004 & RRab & -1.23$\pm$0.31 &                \\                 
 V56   & 13:25:55.44 & -47:37:44.4 & 0.568023   & 13.629 & 0.010 & 13.302 & 0.003 & RRab & -1.26$\pm$0.15 &                \\  
 V57   & 13:27:49.42 & -47:36:50.7 & 0.794402   & 13.190 & 0.003 & 12.846 & 0.002 & RRab & -1.89$\pm$0.14 &                \\                 
 V58   & 13:26:13.03 & -47:24:03.4 & 0.369880   & 13.529 & 0.005 & 13.317 & 0.002 & RRc  & -1.37$\pm$0.18 & -1.91$\pm$0.31 \\                 
 V59   & 13:26:18.41 & -47:29:47.2 & 0.518506   & 13.622 & 0.010 & 13.363 & 0.006 & RRab & -1.00$\pm$0.28 &                \\  
 V62   & 13:26:26.57 & -47:27:55.9 & 0.619770   & 13.330 & 0.018 & 13.062 & 0.004 & RRab & -1.62$\pm$0.29 &                \\                 
 V63   & 13:25:07.87 & -47:36:53.8 & 0.825943   & 13.171 & 0.005 & 12.819 & 0.002 & RRab & -1.73$\pm$0.09 &                \\                 
 V64   & 13:26:02.16 & -47:36:19.6 & 0.344458   & 13.627 & 0.005 & 13.382 & 0.002 & RRc  & -1.46$\pm$0.23 &                \\                 
 V66   & 13:26:33.02 & -47:22:25.5 & 0.407461   & 13.441 & 0.004 & 13.200 & 0.002 & RRc  & -1.68$\pm$0.34 &                \\                 
 V67$^{c}$ & 13:26:28.56 & -47:18:47.2 & 0.564451   & 13.557 & 0.010 & 13.271 & 0.003 & RRab & -1.10          & -1.19$\pm$0.23 \\  
 V68   & 13:26:12.79 & -47:19:36.1 & 0.534696   & 13.175 & 0.005 & 12.896 & 0.002 & RRc  & -1.60$\pm$0.01 &                 \\                       
 V69   & 13:25:10.94 & -47:37:33.2 & 0.653221   & 13.367 & 0.015 & 13.061 & 0.003 & RRab & -1.52$\pm$0.14 &                 \\                       
 V70   & 13:27:27.75 & -47:33:43.3 & 0.390625   & 13.513 & 0.009 & 13.267 & 0.004 & RRc  & -1.94$\pm$0.15 & -1.74$\pm$0.30  \\                      
 V71   & 13:27:08.05 & -47:27:52.2 & 0.357544   & 13.535 & 0.021 & 13.310 & 0.007 & RRc  &                & -1.74$\pm$0.28  \\                      
 V72   & 13:27:33.02 & -47:16:22.7 & 0.384522   & 13.524 & 0.004 & 13.278 & 0.002 & RRc  & -1.32$\pm$0.22 &                 \\                       
 V73   & 13:25:53.65 & -47:16:10.7 & 0.575215   & 13.470 & 0.009 & 13.193 & 0.003 & RRab & -1.50$\pm$0.09 &                 \\ 
 V74   & 13:27:07.25 & -47:17:34.3 & 0.503209   & 13.600 & 0.022 & 13.308 & 0.004 & RRab & -1.83$\pm$0.36 &                 \\  
 V75   & 13:27:19.69 & -47:18:46.9 & 0.421980   & 13.390 & 0.005 & 13.125 & 0.002 & RRc  & -1.49$\pm$0.08 & -1.82$\pm$0.99  \\                        
 V76   & 13:26:57.27 & -47:20:07.9 & 0.337962   & 13.632 & 0.004 & 13.412 & 0.002 & RRc  & -1.45$\pm$0.13 &                 \\                         
 V77$^{c}$ & 13:27:20.87 & -47:22:06.0 & 0.426136   & 13.423 & 0.004 & 13.146 & 0.002 & RRc  & -1.81          & -1.84$\pm$0.43  \\                         	
 V79   & 13:28:25.08 & -47:29:24.8 & 0.608276   & 13.439 & 0.017 & 13.116 & 0.003 & RRab & -1.39$\pm$0.18 &                 \\                         
 V81   & 13:27:36.72 & -47:24:48.4 & 0.389392   & 13.507 & 0.004 & 13.255 & 0.002 & RRc  & -1.72$\pm$0.31 & -1.99$\pm$0.43  \\                         
 V82   & 13:27:35.58 & -47:26:30.8 & 0.335758   & 13.599 & 0.005 & 13.375 & 0.002 & RRc  & -1.56$\pm$0.20 & -1.71$\pm$0.56  \\  
 V83   & 13:27:08.42 & -47:21:34.4 & 0.356612   & 13.594 & 0.004 & 13.359 & 0.002 & RRc  & -1.30$\pm$0.22 &                 \\                                          
 V84   & 13:24:47.47 & -47:29:56.2 & 0.579873   & 13.043 & 0.006 & 12.719 & 0.003 & RRab & -1.47$\pm$0.10 &                 \\                                              
 V85   & 13:25:06.60 & -47:23:33.6 & 0.742758   & 13.245 & 0.010 & 12.907 & 0.003 & RRab & -1.87$\pm$0.31 &                 \\                                      
 \hline                                                                                                                                                          
\end{tabular}                                                                                                                                                    
\end{table*}                                                                                                                                                     
                                                                                                                                                                 
\addtocounter{table}{-1}                                                                                                                                         
                                                                                                                                                                 
\begin{table*}                                                                                                                                                   
\centering                                                                                                                                                       
\caption[]{{\it continued}}                                                                                                                                      
\begin{tabular}{lcccccccccl}                                                                                                                                    
\hline \hline                                                                                                                                                    
  ID & RA (J2000.0)  & DEC (J2000.0) & $P$    & J     & e     & Ks    & e         & Type & [Fe/H]         & [Fe/H]          \\                                                   
     & hh:mm:ss.ss    & dd:mm:ss.s     & (days) & (mag) & (mag) & (mag) & (mag)   &      & (R00)          & (S06)            \\                                                         
\hline                                                                                                                                                 
 V86   & 13:27:15.17 & -47:26:11.6 & 0.647844   & 13.357 & 0.015 & 13.043 & 0.003 & RRab & -1.81$\pm$0.18 & -1.99$\pm$0.23     \\                                         	 
 V87   & 13:26:57.47 & -47:25:35.6 & 0.396488   & 13.479 & 0.007 & 13.246 & 0.003 & RRc  & -1.44$\pm$0.19 &                    \\                                          
 V88   & 13:26:55.89 & -47:25:16.5 & 0.690211   & 13.262 & 0.021 & 12.985 & 0.007 & RRab & -1.65$\pm$0.23 &                    \\                                          
 V89   & 13:26:45.95 & -47:26:01.1 & 0.375110   & 13.574 & 0.017 & 13.325 & 0.005 & RRc  & -1.37$\pm$0.28 & -1.66$\pm$0.23     \\                                        
 V90$^{c}$ & 13:26:45.72 & -47:26:23.6 & 0.603404   & 13.396 & 0.025 & 13.103 & 0.005 & RRab & -2.21          & -1.78$\pm$0.31     \\                                        
 V91   & 13:26:50.58 & -47:26:15.7 & 0.895225   & 13.010 & 0.008 & 12.686 & 0.004 & RRab & -1.44$\pm$0.17 & -1.81$\pm$0.30     \\                                        
 V94   & 13:25:57.07 & -47:22:46.4 & 0.253936   & 13.977 & 0.006 & 13.811 & 0.002 & RRc  & -1.00$\pm$0.11 &                    \\ 
 V95   & 13:25:24.90 & -47:28:52.9 & 0.405067   & 13.502 & 0.005 & 13.223 & 0.002 & RRc  & -1.84$\pm$0.55 &                    \\                                          
 V96$^{c}$ & 13:26:39.27 & -47:27:03.2 & 0.624527   & 13.327 & 0.024 & 13.054 & 0.004 & RRab & -1.22          &                    \\
 V97   & 13:27:08.48 & -47:25:31.4 & 0.691898   & 13.300 & 0.013 & 12.988 & 0.003 & RRab & -1.56$\pm$0.37 & -1.74$\pm$0.17     \\  
 V98   & 13:27:05.83 & -47:26:57.1 & 0.280566   & 13.893 & 0.006 & 13.704 & 0.003 & RRc  & -1.05$\pm$0.12 &                    \\                                          
 V99   & 13:27:02.21 & -47:27:49.6 & 0.766181   & 13.110 & 0.024 & 12.838 & 0.004 & RRab & -1.66$\pm$0.14 & -1.91$\pm$0.25     \\                                           
 V100  & 13:27:04.01 & -47:27:33.8 & 0.552745   & 13.619 & 0.025 & 13.328 & 0.005 & RRab & -1.58$\pm$0.14 &                    \\                                          
 V101  & 13:27:30.22 & -47:29:51.6 & 0.340946   & 13.644 & 0.006 & 13.402 & 0.002 & RRc  & -1.88$\pm$0.32 &                    \\   
 V102  & 13:27:22.08 & -47:30:12.9 & 0.691396   & 13.281 & 0.010 & 12.977 & 0.003 & RRab & -1.84$\pm$0.13 & -1.65$\pm$0.16     \\                                           
 V103  & 13:27:14.26 & -47:28:36.9 & 0.328852   & 13.600 & 0.004 & 13.403 & 0.002 & RRc  & -1.92$\pm$0.11 & -1.78$\pm$0.27     \\                                           
 V104  & 13:28:07.80 & -47:33:44.7 & 0.866308   & 13.168 & 0.006 & 12.824 & 0.003 & RRab & -1.83$\pm$0.18 &                    \\                                          
 V105  & 13:27:46.03 & -47:32:44.3 & 0.335328   & 13.740 & 0.004 & 13.501 & 0.002 & RRc  & -1.24$\pm$0.18 &                    \\                                          
 V106  & 13:26:59.16 & -47:28:12.9 & 0.569903   & 13.362 & 0.025 & 13.148 & 0.005 & RRab & -1.50$\pm$0.23 & -1.90$\pm$0.26     \\  
 V107  & 13:27:14.02 & -47:30:58.5 & 0.514102   & 13.670 & 0.016 & 13.386 & 0.003 & RRab & -1.36$\pm$0.11 &                    \\                                          
 V108  & 13:27:04.66 & -47:29:26.1 & 0.594458   & 13.328 & 0.011 & 13.078 & 0.004 & RRab & -1.93$\pm$0.23 & -1.63$\pm$0.13     \\                                           
 V109  & 13:27:01.52 & -47:29:37.0 & 0.744098   & 13.193 & 0.020 & 12.907 & 0.003 & RRab & -1.51$\pm$0.25 & -1.70$\pm$0.07     \\                                           
 V110  & 13:27:02.04 & -47:30:07.1 & 0.332107   & 13.641 & 0.008 & 13.451 & 0.003 & RRc  & -2.14$\pm$0.16 & -1.65$\pm$0.52     \\   
 V111  & 13:26:48.99 & -47:28:40.5 & 0.762905   & 13.184 & 0.014 & 12.866 & 0.003 & RRab & -1.66$\pm$0.04 & -1.79$\pm$0.09     \\                                           
 V112  & 13:26:54.24 & -47:30:23.6 & 0.474359   & 13.574 & 0.027 & 13.354 & 0.006 & RRab & -1.81$\pm$0.26 &                    \\ 
 V113  & 13:26:56.29 & -47:31:47.9 & 0.573375   & 13.447 & 0.025 & 13.192 & 0.005 & RRab & -1.65$\pm$0.34 &                    \\                                           
 V114  & 13:26:50.10 & -47:30:21.4 & 0.675307   & 13.292 & 0.015 & 12.997 & 0.004 & RRab & -1.32$\pm$0.30 & -1.61$\pm$0.99     \\                                         
 V115  & 13:26:12.27 & -47:34:17.9 & 0.630474   & 13.389 & 0.015 & 13.095 & 0.004 & RRab & -1.87$\pm$0.01 & -1.64$\pm$0.32     \\ 
 V116  & 13:26:35.47 & -47:28:07.2 & 0.720133   & 13.231 & 0.018 & 12.980 & 0.006 & RRab & -1.27$\pm$0.44 & -1.11$\pm$0.17     \\                                         
 V117  & 13:26:19.88 & -47:29:21.6 & 0.421641   & 13.439 & 0.005 & 13.191 & 0.003 & RRc  & -1.68$\pm$0.25 &                    \\                                          
 V119  & 13:26:38.27 & -47:31:18.3 & 0.305876   & 13.695 & 0.004 & 13.516 & 0.002 & RRc  & -1.61$\pm$0.10 &                    \\  
 V120  & 13:26:25.51 & -47:32:49.0 & 0.548537   & 13.560 & 0.021 & 13.291 & 0.005 & RRab & -1.39$\pm$0.06 & -1.15$\pm$0.16     \\  
 V121  & 13:26:28.15 & -47:31:51.0 & 0.304182   & 13.636 & 0.004 & 13.455 & 0.002 & RRc  & -1.46$\pm$0.13 & -1.83$\pm$0.40     \\  
 V122  & 13:26:30.30 & -47:33:02.5 & 0.634929   & 13.340 & 0.020 & 13.062 & 0.004 & RRab & -2.02$\pm$0.18 & -1.79$\pm$0.21     \\                                             
 V123  & 13:26:51.07 & -47:37:13.3 & 0.473884   & 13.389 & 0.005 & 13.110 & 0.002 & RRc  & -1.64$\pm$0.01 &                    \\  
 V124  & 13:26:54.38 & -47:39:07.6 & 0.331860   & 13.650 & 0.005 & 13.433 & 0.002 & RRc  & -1.33$\pm$0.23 &                    \\                                           
 V125  & 13:26:48.98 & -47:41:03.7 & 0.592888   & 13.416 & 0.011 & 13.145 & 0.003 & RRab & -1.67$\pm$0.22 & -1.81$\pm$0.38     \\                                          
 V126  & 13:28:08.14 & -47:40:46.5 & 0.341891   & 13.646 & 0.006 & 13.403 & 0.002 & RRc  & -1.31$\pm$0.13 &                    \\                                          
 V127  & 13:25:19.43 & -47:28:37.6 & 0.305274   & 13.738 & 0.004 & 13.518 & 0.002 & RRc  & -1.59$\pm$0.08 &                    \\  
 V128  & 13:26:17.72 & -47:30:13.5 & 0.834988   & 13.076 & 0.008 & 12.761 & 0.003 & RRab & -1.88$\pm$0.04 &                    \\                                          
 V130  & 13:26:10.00 & -47:13:40.0 & 0.493250   & 13.675 & 0.008 & 13.406 & 0.003 & RRab & -1.46$\pm$0.17 &                   \\  
 V131  & 13:26:30.04 & -47:29:41.1 & 0.392123   & 13.402 & 0.006 & 13.180 & 0.003 & RRc  & -1.56$\pm$0.20 & -1.66$\pm$0.48    \\                                            
 V132  & 13:26:39.18 & -47:29:10.0 & 0.655656   & 13.267 & 0.022 & 12.990 & 0.005 & RRab & -1.91$\pm$0.20 &                   \\                                          
 V134  & 13:25:13.33 & -47:12:28.4 & 0.652903   & 13.339 & 0.018 & 13.040 & 0.006 & RRab & -1.80$\pm$0.41 &                   \\  
 V135$^{a}$ & 13:26:28.06 & -47:29:18.3 & 0.632527   & 12.877 & 0.029 & 12.536 & 0.011 & RRab & -2.20          & -1.57$\pm$0.18   \\
 V136  & 13:26:31.06 & -47:27:40.9 & 0.391945   & 13.397 & 0.007 & 13.182 & 0.005 & RRc  & -1.83$\pm$0.47 & -1.64$\pm$0.37    \\  
 V137  & 13:26:31.52 & -47:27:04.6 & 0.334205   & 13.557 & 0.007 & 13.352 & 0.002 & RRc  & -1.19$\pm$0.18 &                   \\                                          
 V139$^{a}$ & 13:26:37.75 & -47:27:35.4 & 0.676871   & 13.022 & 0.028 & 12.707 & 0.008 & RRab & -1.46$\pm$0.04 & -1.83$\pm$0.20   \\
 V140  & 13:26:42.15 & -47:30:07.5 & 0.619849   & 13.355 & 0.019 & 13.087 & 0.006 & RRab &                & -1.72$\pm$0.15    \\  
 V141  & 13:26:40.87 & -47:29:28.2 & 0.697363   & 13.221 & 0.019 & 12.915 & 0.007 & RRab & -1.55$\pm$0.36 & -2.20$\pm$0.36    \\  
 V142  & 13:26:42.63 & -47:28:42.9 & 0.375877   & 13.520 & 0.014 & 13.284 & 0.006 & RRc  &                & -1.81$\pm$0.24    \\  
 V144  & 13:26:43.02 & -47:28:18.0 & 0.835320   & 13.044 & 0.008 & 12.742 & 0.004 & RRab & -1.71$\pm$0.12 &                   \\                                                   
 V145  & 13:26:51.21 & -47:31:08.8 & 0.373214   & 13.542 & 0.005 & 13.317 & 0.002 & RRc  & -1.58$\pm$0.07 &                   \\  
 V147  & 13:27:15.88 & -47:31:10.3 & 0.422615   & 13.387 & 0.006 & 13.166 & 0.002 & RRc  & -1.66$\pm$0.14 &                   \\  
 V149  & 13:27:32.85 & -47:13:43.3 & 0.682728   & 13.297 & 0.013 & 12.997 & 0.003 & RRab & -1.21$\pm$0.24 &                   \\                                                       
 V150  & 13:27:40.23 & -47:36:00.4 & 0.899302   & 13.086 & 0.008 & 12.766 & 0.003 & RRab & -1.76$\pm$0.34 &                   \\                                                       
 V151  & 13:28:25.31 & -47:16:00.0 & 0.407756   & 13.470 & 0.003 & 13.213 & 0.002 & RRc  & -1.30$\pm$0.24 &                   \\                                 
 V153  & 13:26:49.65 & -47:26:23.8 & 0.386245   & 13.511 & 0.006 & 13.281 & 0.003 & RRc  & -1.38$\pm$0.19 &                   \\                            
 V154  & 13:27:03.11 & -47:30:32.4 & 0.322340   & 13.618 & 0.007 & 13.462 & 0.002 & RRc  & -1.39$\pm$0.12 & -1.49$\pm$0.23    \\                                    
 V155  & 13:26:53.64 & -47:24:42.8 & 0.413925   & 13.428 & 0.006 & 13.187 & 0.003 & RRc  & -1.46$\pm$0.09 &                   \\                                    
 V156  & 13:26:47.87 & -47:31:52.6 & 0.359067   & 13.513 & 0.009 & 13.320 & 0.004 & RRc  & -1.40$\pm$0.04 & -1.51$\pm$0.38    \\                                  
 V157  & 13:26:46.45 & -47:27:17.7 & 0.406578   & 13.493 & 0.006 & 13.260 & 0.003 & RRc  & -1.49$\pm$0.10 &                   \\                                    
 V158  & 13:26:45.30 & -47:30:40.4 & 0.367276   & 13.543 & 0.009 & 13.336 & 0.003 & RRc  & -1.25$\pm$0.06 & -1.64$\pm$0.49    \\                                 
 V160$^{c}$ & 13:25:36.08 & -47:12:32.3 & 0.397527   & 13.497 & 0.004 & 13.237 & 0.002 & RRc  & -1.66          &                   \\
 V163  & 13:25:49.47 & -47:20:21.8 & 0.313196   & 13.701 & 0.005 & 13.493 & 0.002 & RRc  & -1.18$\pm$0.27 &                   \\                                  
 V169  & 13:27:20.45 & -47:23:59.6 & 0.319116   & 13.713 & 0.003 & 13.502 & 0.001 & RRc  &                & -1.65$\pm$0.19    \\  
 V261$^{a}$  & 13:27:15.40 & -47:21:29.9 & 0.402512   & 13.358 & 0.005 & 13.049 & 0.002 & RRc &                & -1.50$\pm$0.35 \\
  \hline                                                                                                                                                                          
\end{tabular}                                                                                                                                                                     
\end{table*}                                                                                                                                                                       
                                                                                                                                                                                        
\addtocounter{table}{-1}                                                                                                                                                                    
                                                                                                                                                                                            
\begin{table*}                                                                                                                                                                                
\centering                                                                                                                                                                                
\caption[]{{\it continued}}                                                                                                                                                               
\begin{tabular}{lcccccccccl}
\hline \hline                                                                                                                                                                               
  ID & RA (J2000.0)  & DEC (J2000.0) & $P$    & J     & e     & Ks    & e     & Type     & [Fe/H] & [Fe/H]         \\                                                                                       
     & hh:mm:ss.ss    & dd:mm:ss.s     & (days) & (mag) & (mag) & (mag) & (mag) &        & (R00)  & (S06)          \\                                                                                        
\hline                                                                                                                                                                             
 V263  & 13:26:13.11 & -47:26:10.2 & 1.012158   & 12.970 & 0.004 & 12.625 & 0.002 & RRab &        & -1.73$\pm$0.19  \\                                               
 V265  & 13:26:30.19 & -47:28:45.6 & 0.422600   & 13.384 & 0.02  & 13.129 & 0.004 & RRc  &        & -2.00$\pm$0.29  \\                                                
 V267  & 13:26:40.18 & -47:26:36.0 & 0.315822   & 13.624 & 0.008 & 13.456 & 0.002 & RRc  &        & -1.62$\pm$0.63  \\ 
 V268  & 13:26:35.11 & -47:26:11.2 & 0.8129220  & 13.145 & 0.006 & 12.822 & 0.002 & RRab &        & -1.76$\pm$0.24  \\                                             
 V271  & 13:26:47.10 & -47:30:04.3 & 0.4432000  & 13.337 & 0.008 & 13.100 & 0.003 & RRc  &        & -1.80$\pm$0.21  \\                                                    	
 V275  & 13:26:49.72 & -47:27:37.4 & 0.3778970  & 13.526 & 0.009 & 13.314 & 0.003 & RRc  &        & -1.66$\pm$0.36  \\                                                    	
 V341  & 13:26:54.63 & -47:28:48.4 & 0.3061360  & 13.621 & 0.017 & 13.366 & 0.015 & RRc  &        & -1.78$\pm$0.59  \\                                                    
 V342  & 13:27:18.72 & -47:28:22.6 & 0.3083890  & 13.687 & 0.005 & 13.485 & 0.002 & RRc  &        & -1.71$\pm$0.55   \\                                                   
 V346  & 13:26:46.91 & -47:28:14.3 & 0.3276230  & 13.581 & 0.012 & 13.402 & 0.004 & RRc  &        & -1.52$\pm$0.54   \\                                                    
 V347  & 13:26:50.82 & -47:27:46.2 & 0.3288490  & 13.637 & 0.025 & 13.459 & 0.009 & RRc  &        & -1.66$\pm$0.27   \\                                                    
 V350  & 13:26:56.44 & -47:30:50.2 & 0.3791080  & 13.442 & 0.012 & 13.236 & 0.005 & RRc  &        & -1.45$\pm$0.40  \\                                                    	
 V353$^{a}$ & 13:26:43.61 & -47:27:57.6 & 0.4010200  & 13.245 & 0.026 & 12.977 & 0.014 & RRc  &        & -1.93$\pm$0.31  \\
 V354  & 13:26:38.58 & -47:25:10.2 & 0.4200900  & 13.419 & 0.009 & 13.182 & 0.004 & RRc  &        & -1.73$\pm$0.23 \\                                                   
 V357  & 13:26:17.73 & -47:30:23.0 & 0.2977750  & 13.690 & 0.005 & 13.510 & 0.003 & RRc  &        & -1.64$\pm$0.99 \\   
 V366$^{b}$ & 13:26:41.54 & -47:31:42.2 & 0.9999100  & 12.748 & 0.007 & 12.423 & 0.004 & RRab &        & -1.61$\pm$0.14 \\
 V399  & 13:26:29.51 & -47:30:03.0 & 0.3097820  & 13.659 & 0.007 & 13.495 & 0.003 & RRc  &        & -1.70$\pm$0.67 \\                                                     
\hline
\end{tabular}
\begin{flushleft}
\footnotesize{$^{a}$ Stars not considered to derive the PL-$Z$ relations because their mean magnitudes are brighter than the mean locus of RRab or RRc stars (see Figure~\ref{plrrl}). These ``over-luminous'' stars are due to unresolved companions and blending. \\ $^{b}$ The $J$ and $K_{\rm S}$ light curves have a considerable gap in the observations because of the one day period. This star was not considered to derive the PL-$Z$ relations. \\ $^{c}$ Stars with metallicity measurements without error reported were not included to derive the PL-$Z$ relations.}
\end{flushleft}
\end{table*}

\begin{table*}[ht!]
\centering
\caption[]{Catalogue of fundamental-mode candidate SX Phe stars in $\omega$~Cen, used to derive the empirical PL relations. \label{tab:sx}}
\begin{tabular}{lcccccccl}
\hline \hline
  ID       & RA (J2000.0)  & DEC (J2000.0) & $P$    & J     & e     & Ks    & e       & Remarks\tablefootmark{a}\\
           & hh:mm:ss.ss    & dd:mm:ss.s     & (days) & (mag) & (mag) & (mag) & (mag) &  \\
\hline           
 V194      & 13:27:53.95 & -47:31:54.2 & 0.0471777  & 16.237 & 0.007 & 16.106 & 0.006 & 1         \\                                   
 V195      & 13:27:15.63 & -47:24:34.9 & 0.0654912  & 15.808 & 0.006 & 15.591 & 0.005 & 1          \\                                   
 V196      & 13:25:01.21 & -47:25:29.6 & 0.0574000  & 16.080 & 0.009 & 15.858 & 0.010 & 1          \\                                   
 V197      & 13:26:20.40 & -47:31:59.7 & 0.0471210  & 16.153 & 0.044 & 15.980 & 0.067 & 2  \\                                
 V198      & 13:26:34.53 & -47:31:03.7 & 0.0481817  & 16.348 & 0.044 & 16.212 & 0.058 & 2  \\ 
 V199      & 13:26:28.63 & -47:28:38.4 & 0.0622866  & 15.669 & 0.007 & 15.468 & 0.004 & 1          \\ 
 V201      & 13:26:11.09 & -47:15:54.1 & 0.0506500  & 16.250 & 0.029 & 16.061 & 0.054 & 2  \\ 
 V202      & 13:26:38.97 & -47:11:51.2 & 0.0464200  & 16.319 & 0.018 & 16.125 & 0.037 & 2     \\
 V204      & 13:27:07.90 & -47:37:05.4 & 0.0493757  & 16.110 & 0.008 & 15.960 & 0.008 & 1              \\
 V217      & 13:26:16.91 & -47:27:25.9 & 0.0532600  & 16.076 & 0.029 & 15.859 & 0.037 & 2      \\ 
 V218      & 13:26:11.22 & -47:17:54.1 & 0.0437393  & 16.335 & 0.019 & 16.177 & 0.035 & 2      \\ 
 V219      & 13:26:08.43 & -47:19:24.6 & 0.0386681  & 16.544 & 0.022 & 16.455 & 0.048 & 2      \\ 
 V220      & 13:26:48.61 & -47:21:42.4 & 0.0528868  & 16.127 & 0.007 & 15.936 & 0.006 & 1      \\
 V222      & 13:26:18.52 & -47:41:12.5 & 0.0389100  & 16.475 & 0.022 & 16.350 & 0.039 & 2      \\
 V225      & 13:27:02.28 & -47:24:36.8 & 0.0486381  & 16.055 & 0.011 & 15.918 & 0.008 & 1     \\
 V226      & 13:26:10.47 & -47:29:57.3 & 0.0378524  & 16.529 & 0.027 & 16.422 & 0.049 & 2      \\
 V227      & 13:26:14.36 & -47:23:54.2 & 0.0382255  & 16.508 & 0.027 & 16.346 & 0.047 & 2      \\
 V228      & 13:26:40.48 & -47:33:46.6 & 0.0398531  & 16.456 & 0.027 & 16.347 & 0.056 & 2      \\
 V229      & 13:27:12.60 & -47:23:58.9 & 0.0375333  & 16.524 & 0.023 & 16.351 & 0.039 & 2      \\
 V231      & 13:27:19.03 & -47:35:54.4 & 0.0374850  & 16.601 & 0.024 & 16.479 & 0.048 & 2      \\
 V232      & 13:26:17.32 & -47:11:49.9 & 0.0369700  & 16.647 & 0.022 & 16.506 & 0.042 & 2      \\
 V233      & 13:26:47.28 & -47:20:32.0 & 0.0365377  & 16.478 & 0.027 & 16.386 & 0.056 & 2      \\
 V237      & 13:25:52.88 & -47:23:56.0 & 0.0656024  & 15.905 & 0.007 & 15.675 & 0.004 & 1       \\
 V238      & 13:27:33.96 & -47:29:23.4 & 0.0408004  & 16.484 & 0.021 & 16.372 & 0.044 & 2     \\
 V249      & 13:25:46.27 & -47:26:28.8 & 0.0349468  & 16.624 & 0.025 & 16.425 & 0.041 & 2     \\ 
 V250      & 13:27:31.74 & -47:35:42.1 & 0.0406269  & 16.548 & 0.024 & 16.388 & 0.045 & 2     \\
 V252      & 13:27:29.71 & -47:29:00.5 & 0.0466226  & 16.398 & 0.019 & 16.178 & 0.031 & 2     \\
 V253      & 13:27:22.13 & -47:27:52.3 & 0.0399687  & 16.421 & 0.023 & 16.285 & 0.044 & 2     \\
 V260      & 13:24:52.73 & -47:25:22.0 & 0.0462600  & 16.271 & 0.018 & 16.073 & 0.028 & 2     \\
 V299      & 13:27:00.30 & -47:22:29.9 & 0.0344409  & 16.654 & 0.034 & 16.477 & 0.063 & 2  \\
 V300      & 13:25:46.55 & -47:24:00.7 & 0.0347301  & 16.640 & 0.023 & 16.437 & 0.040 & 2  \\ 
 V304      & 13:26:19.57 & -47:20:01.6 & 0.0361405  & 16.488 & 0.028 & 16.314 & 0.045 & 2  \\
 V305      & 13:27:51.42 & -47:19:51.4 & 0.0365673  & 16.474 & 0.022 & 16.266 & 0.035 & 2  \\
 V306      & 13:27:10.99 & -47:28:24.8 & 0.0384044  & 16.551 & 0.033 & 16.364 & 0.069 & 2  \\
 V308      & 13:27:50.03 & -47:21:16.1 & 0.0389852  & 16.480 & 0.022 & 16.324 & 0.037 & 2  \\
 V311      & 13:26:58.75 & -47:29:49.9 & 0.0414133  & 16.295 & 0.078 & 15.976 & 0.118 & 2  \\ 
 V313      & 13:26:38.28 & -47:31:37.3 & 0.0418484  & 16.443 & 0.034 & 16.277 & 0.065 & 2 \\
 V316      & 13:26:14.84 & -47:31:10.8 & 0.0424040  & 16.405 & 0.025 & 16.264 & 0.041 & 2  \\
 V319      & 13:26:28.52 & -47:31:02.9 & 0.0489421  & 16.304 & 0.028 & 16.139 & 0.056 & 2 \\
 V320      & 13:26:49.43 & -47:34:29.1 & 0.0471934  & 16.351 & 0.028 & 16.181 & 0.038 & 2  \\
 V322      & 13:26:38.85 & -47:27:39.1 & 0.0479562  & 16.009 & 0.037 & 15.853 & 0.052 & 2  \\
 V326      & 13:26:40.17 & -47:24:55.7 & 0.0569058  & 15.994 & 0.026 & 15.817 & 0.035 & 2  \\
 V419/W22  & 13:28:53.57 & -47:19:29.7 & 0.0410000  & 16.403 & 0.020 & 16.236 & 0.032 & 2 \\
 V420/W23  & 13:28:44.57 & -47:24:51.0 & 0.0370000  & 16.472 & 0.021 & 16.322 & 0.036 & 2 \\ 
 V445/W131 & 13:24:54.09 & -47:41:03.5 & 0.0470000  & 16.350 & 0.020 & 16.146 & 0.030 & 2  \\ 
\hline
 \hline                                                                                                                                                          
\end{tabular}     
\begin{flushleft}
\footnotesize{$^{a}$1: Variability recovered in our data; 2: Variability not recovered in our data.} 
\end{flushleft}
\end{table*}                       

\clearpage
  
\end{appendix}

\end{document}